\documentclass[12pt]{article}
\title{Partitioning schemes for quicksort and quickselect}
\author{Krzysztof C. Kiwiel\thanks{Systems Research Institute,
        Newelska 6, 01--447 Warsaw, Poland
        ({\tt kiwiel@ibspan.waw.pl})}}
\date{December 20, 2003}

\newcommand{\BbbF}{{\rm\normalcolor I\kern-.18em F}}
\newcommand{\BbbR}{{\rm\normalcolor I\kern-.18em R}}
\newcommand{\eqref}[1]{{\normalfont\normalcolor(\ref{#1})}}
\makeatletter
\def\proof{%
   \def\a##1{\begin{trivlist}\item[]{\bf\ignorespaces{##1}.}%
    \enspace\ignorespaces}%
   \def\b[##1]{\a{Proof\ \ignorespaces{##1}}}%
   \@ifnextchar[{\b}{\a{Proof}}}
\def\endproof{\end{trivlist}}
\def\qed{\relax\protect\ifmmode\ifinner\else\quad\fi\fi
    \hbox{\vbox{\hrule height.4pt\hbox{\vbox{\hrule height.4pt
    \hbox{\vrule width.4pt\vphantom{\normalsize A}\kern.5em
    \vrule width.4pt}\hrule height.4pt}}}}}
\newtoks\@stequation
\def\subequations{\refstepcounter{equation}%
\edef\@savedequation{\the\c@equation}%
\@stequation=\expandafter{\theequation}
\edef\@savedtheequation{\the\@stequation}
\edef\oldtheequation{\theequation}%
\setcounter{equation}{0}%
\def\theequation{\oldtheequation\alph{equation}}}%
\def\endsubequations{%
\setcounter{equation}{\@savedequation}%
\@stequation=\expandafter{\@savedtheequation}%
\edef\theequation{\the\@stequation}\global\@ignoretrue}
\def\@begintheorem#1#2{\trivlist
    \item[\hskip \labelsep{\bfseries #1\ #2.}]\itshape}
\def\@opargbegintheorem#1#2#3{\trivlist
    \item[\hskip \labelsep{\bfseries #1\ #2\ (#3).}]\itshape}
\@addtoreset{equation}{section}
\def\theequation{\thesection.\arabic{equation}}
\@addtoreset{figure}{section}

\@addtoreset{table}{section}

\let\@@eqnsel=\relax
\def\@tempa{%
    \stepcounter{equation}%
    \def\@currentlabel{\p@equation\theequation}%
    \global\@eqnswtrue\m@th
    \global\@eqcnt\z@
    \tabskip\mathindent
    \let\\=\@eqncr
    \setlength\abovedisplayskip{\topsep}%
    \ifvmode
      \addtolength\abovedisplayskip{\partopsep}%
    \fi
    \addtolength\abovedisplayskip{\parskip}%
    \setlength\belowdisplayskip{\abovedisplayskip}%
    \setlength\belowdisplayshortskip{\abovedisplayskip}%
    \setlength\abovedisplayshortskip{\abovedisplayskip}%
    $$\everycr{}\halign to\linewidth
    \bgroup
      \hskip\@centering
      $\displaystyle\tabskip\z@skip{##}$\@eqnsel&%
      \global\@eqcnt\@ne \hskip \tw@\arraycolsep \hfil${##}$\hfil&%
      \global\@eqcnt\tw@ \hskip \tw@\arraycolsep
        $\displaystyle{##}$\hfil \tabskip\@centering&%
      \global\@eqcnt\thr@@
        \hb@xt@\z@\bgroup\hss##\egroup\tabskip\z@skip\cr}%
\def\@tempb{%
   \stepcounter{equation}%
   \def\@currentlabel{\p@equation\theequation}%
   \global\@eqnswtrue
   \m@th
   \global\@eqcnt\z@
   \tabskip\@centering
   \let\\\@eqncr
   $$\everycr{}\halign to\displaywidth\bgroup
       \hskip\@centering$\displaystyle\tabskip\z@skip{##}$\@eqnsel
      &\global\@eqcnt\@ne\hskip \tw@\arraycolsep \hfil${##}$\hfil
      &\global\@eqcnt\tw@ \hskip \tw@\arraycolsep
         $\displaystyle{##}$\hfil\tabskip\@centering
      &\global\@eqcnt\thr@@ \hb@xt@\z@\bgroup\hss##\egroup
         \tabskip\z@skip
      \cr
}
\ifx\eqnarray\@tempa
    \def\eqnarray{%
    \stepcounter{equation}%
    \def\@currentlabel{\p@equation\theequation}%
    \global\@eqnswtrue\m@th
    \global\@eqcnt\z@
    \tabskip\mathindent
    \let\\=\@eqncr
    \setlength\abovedisplayskip{\topsep}%
    \ifvmode
      \addtolength\abovedisplayskip{\partopsep}%
    \fi
    \addtolength\abovedisplayskip{\parskip}%
    \setlength\belowdisplayskip{\abovedisplayskip}%
    \setlength\belowdisplayshortskip{\abovedisplayskip}%
    \setlength\abovedisplayshortskip{\abovedisplayskip}%
    $$\everycr{}\halign to\linewidth
    \bgroup
      \hskip\@centering
      $\displaystyle\tabskip\z@skip{##}$\@eqnsel&%
      \global\@eqcnt\@ne
      \@@eqnsel
      \hfil${{}##{}}$\hfil&
      \global\@eqcnt\tw@
      \@@eqnsel
        $\displaystyle{##}$\hfil \tabskip\@centering&%
      \global\@eqcnt\thr@@
        \hb@xt@\z@\bgroup\hss##\egroup\tabskip\z@skip\cr}%
\else\ifx\eqnarray\@tempb
   \def\eqnarray{%
   \stepcounter{equation}%
   \def\@currentlabel{\p@equation\theequation}%
   \global\@eqnswtrue
   \m@th
   \global\@eqcnt\z@
   \tabskip\@centering
   \let\\\@eqncr
   $$\everycr{}\halign to\displaywidth\bgroup
       \hskip\@centering$\displaystyle\tabskip\z@skip{##}$\@eqnsel
      &\global\@eqcnt\@ne
      \@@eqnsel
      \hfil${{}##{}}$\hfil
      &\global\@eqcnt\tw@
      \@@eqnsel
         $\displaystyle{##}$\hfil\tabskip\@centering
      &\global\@eqcnt\thr@@ \hb@xt@\z@\bgroup\hss##\egroup
         \tabskip\z@skip
      \cr}
\else 
\fi\fi
\def\@tempa{}			
\def\@tempb{}
\@ifundefined{chapter}{%
  \renewenvironment{thebibliography}[1]
     {\section*{\refname
        \@mkboth{\MakeUppercase\refname}{\MakeUppercase\refname}}%
      \list{\@biblabel{\@arabic\c@enumiv}}%
           {\settowidth\labelwidth{\@biblabel{#1}}%
            \leftmargin\labelwidth
            \advance\leftmargin\labelsep
            \itemsep \z@                 
            \@openbib@code
            \usecounter{enumiv}%
            \let\p@enumiv\@empty
            \renewcommand\theenumiv{\@arabic\c@enumiv}}%
      \sloppy
      \clubpenalty4000
      \@clubpenalty \clubpenalty
      \widowpenalty4000%
      \sfcode`\.\@m}
     {\def\@noitemerr
       {\@latex@warning{Empty `thebibliography' environment}}%
      \endlist}}%
{\renewenvironment{thebibliography}[1]
     {\section*{\bibname
        \@mkboth{\MakeUppercase\bibname}{\MakeUppercase\bibname}}%
      \list{\@biblabel{\@arabic\c@enumiv}}%
           {\settowidth\labelwidth{\@biblabel{#1}}%
            \leftmargin\labelwidth
            \advance\leftmargin\labelsep
            \itemsep \z@                 
            \@openbib@code
            \usecounter{enumiv}%
            \let\p@enumiv\@empty
            \renewcommand\theenumiv{\@arabic\c@enumiv}}%
      \sloppy
      \clubpenalty4000
      \@clubpenalty \clubpenalty
      \widowpenalty4000%
      \sfcode`\.\@m}
     {\def\@noitemerr
       {\@latex@warning{Empty `thebibliography' environment}}%
      \endlist}}%
\newcommand{\Argmax}{{\operator@font Arg}\max}
\newcommand{\Argmin}{{\operator@font Arg}\min}
\newcommand{\argmax}{{\operator@font arg}\max}
\newcommand{\argmin}{{\operator@font arg}\min}
\newcommand{\Exp}{\mathord{\operator@font E}}
\newcommand{\med}{\mathop{\operator@font med}}
\newcommand{\Prob}{\mathord{\operator@font P}}
\newcommand{\rank}{\mathop{\operator@font rank}}
\newcommand{\var}{\mathop{\operator@font var}}
\makeatother
\newtheorem{theorem}{Theorem}[section]
\newtheorem{algorithm}[theorem]{Algorithm}

\newtheorem{corollary}[theorem]{Corollary}

\newtheorem{fact}[theorem]{Fact}
\newtheorem{lemma}[theorem]{Lemma}

\newtheorem{scheme}{Scheme}

\begin{document}           

\maketitle                 

\begin{abstract}
\noindent
We introduce several modifications of the partitioning schemes used
in Hoare's quicksort and quickselect algorithms, including ternary
schemes which identify keys less or greater than the pivot.  We give
estimates for the numbers of swaps made by each scheme.  Our
computational experiments indicate that ternary schemes allow
quickselect to identify all keys equal to the selected key at little
additional cost.
\end{abstract}

\begin{quotation}
\end{quotation}



\section{Introduction}
\label{s:intro}
Hoare's quicksort \cite{hoa:q} and quickselect (originally called
{\sc Find}) \cite{hoa:a65} are among the most widely used algorithms
for sorting and selection.  In our context, given an array
$x[1\colon n]$ of $n$ elements and a total order $<$, sorting means
permuting the elements so that $x_i\le x_{i+1}$ for $i=1\colon n-1$,
whereas for the simpler problem of selecting the $k$th smallest
element, the elements are permuted so that $x_i\le x_k\le x_j$ for
$1\le i\le k\le j\le n$.

Both algorithms choose a pivot element, say $v$, and partition the
input into a left array $x[1\colon a-1]\le v$, a middle array
$x[a\colon b]=v$, and a right array $x[b+1\colon n]\ge v$.  Then
quicksort is called recursively on the left and right arrays, whereas
quickselect is called on the left array if $k<a$, or the right array
if $k>b$; if $a\le k\le b$, selection is finished.

This paper introduces useful modifications of several partitioning
schemes.  First, we show that after exchanging $x_1$ with $x_n$ when
necessary, the classic scheme of Sedgewick
\cite[\S5.2.2]{knu:acpIII2} no longer needs an artificial sentinel.
Second, it turns out that a simple modification of another popular
scheme of Sedgewick \cite[Prog.\ 3]{bemc:esf} allows it to handle
equal keys more efficiently; both schemes take $n$ or $n+1$
comparisons.  Third, we describe a scheme which makes just the $n-1$
necessary comparisons, as well as the minimum number of swaps when the
elements are distinct.  This should be contrasted with Lomuto's scheme
\cite[Prog.\ 2]{bemc:esf}, \cite[\S7.1]{colerist:ia2}, which takes
$n-1$ comparisons but up to $n-1$ swaps.  Hence we analyze the
average numbers of swaps made by the four schemes when the elements
are distinct and in random order.  The first three schemes take at
most $n/4$ swaps on average, whereas Lomuto's scheme takes up to $n-1$.
Further, for the pivot selected as the median of a sample of $2t+1$
elements, the first three schemes make asymptotically $n/6$ swaps for
$t=0$, $n/5$ for $t=1$, etc.\ (cf.\ \S\ref{ss:pivsamplefixed}), while
Lomuto's scheme takes $(n-1)/2$; the swap counts are similar when the pivot
is Tukey's ninther \cite{bemc:esf,chhwts:atc,dur:aao}.

When equal keys occur, one may prefer a {\em ternary\/} scheme which
produces a left array with keys $<v$ and a right array with keys $>v$,
instead of $\le v$ and $\ge v$ as do {\em binary\/} schemes.  Here
only the Bentley-McIlroy scheme \cite{bemc:esf} looks competitive,
since Dijkstra's ``Dutch national flag'' scheme
\cite[Chap.\ 14]{dij:dp} and Wegner's schemes \cite{weg:qek} are more
complex.  However, the four schemes discussed above also have
attractive ternary versions.  Our first scheme omits pointer tests
in its key comparison loops, keeping them as fast as possible.  Our
second scheme improves on another scheme of Sedgewick
\cite[Chap.\ 7, quicksort]{sed:ac++3} (which needn't produce true
ternary partitions; cf.\ \S\ref{ss:stind1}).  Our third scheme is
a simple modification of the Bentley-McIlroy scheme which makes
$n-1$ comparisons; the original version takes $n-1/2$ on average
(cf.\ Lem.\ \ref{l:BMcmp}), although $n-1$ was assumed in
\cite{dur:aao}.  Ternary versions of Lomuto's scheme seem to be less
attractive.  When many equal keys occur, the Bentley-McIlroy scheme
tends to make fewer swaps than the other schemes, but it may swap
needlessly equal keys with themselves and its inner loops involve
pointer tests.  Hence we introduce hybrid two-phase versions which
eliminate vacuous swaps in the first phase and pointer tests in the
second phase.

Ternary schemes, although slower than their simpler binary
counterparts, have at least two advantages.  First, quicksort's
recursive calls aren't made on the equal keys isolated by
partitioning.  Second, quickselect can identify {\em all\/} keys equal
to the $k$th smallest by finding two indices $k_-\le k\le k_+$ such
that $x[1\colon k_--1]< x_k = x[k_-\colon k_+]<x[k_++1\colon n]$ on
output.

Our fairly extensive computational tests with quickselect (we left
quicksort for future work) were quite suprising.  First, the inclusion
of pointer tests in the key comparison loops didn't result in
significant slowdowns; this is in sharp contrast with traditional
recommendations \cite[Ex.\ 5.2.2-24]{knu:acpIII2},
\cite[p.\ 848]{sed:iqp}, but agrees with the observation of
\cite{bemc:esf} that Knuth's MIX cost model needn't be appropriate
for modern machines.  Second, the overheads of ternary schemes
relative to binary schemes were quite mild.  Third, Lomuto's binary
scheme was hopeless when many equal keys occured, since its running
time may be quadratic in the number of keys equal to the $k$th smallest.

More information on theoretical and practical aspects of quicksort
and quickselect can be found in
\cite{bese:fas,gru:mvh,hwts:qdf,kimapr:ahf,maro:oss,mus:iss,val:iss}
and references therein.


The paper is organized as follows.
The four bipartitioning schemes of interest are described in
\S\ref{s:binpart} and their average-case analysis is given in
\S\ref{s:averbi}.  In \S\ref{s:sampsent} we present tuned versions
(cf.\ \cite[\S7]{maro:oss}) for the case where the pivot is selected
from a sample of several elements.  Tripartitioning schemes are
discussed in \S\ref{s:terpart}.
Finally, our computational results are reported in \S\ref{s:exp}.
%
%
\section{Bipartitioning schemes}
\label{s:binpart}
Each invocation of quicksort and quickselect deals with a subarray
$x[l\colon r]$ of the input array $x[1\colon n]$; abusing notation,
we let $n:=r-l+1$ denote the size of the current subarray.  It is
convenient to assume that the pivot $v:=x_l$ is placed first
(after a possible exchange with another element).  Each {\em binary\/}
scheme given below partitions the array into three blocks, with
$x_m\le v$ for $l\le m<a$, $x_m=v$ for $a\le m\le b$, $x_m\ge v$
for $b<m\le r$, $l\le a\le b\le r$.  We suppose that $n>1$ (otherwise
partitioning is trivial: set $a:=b:=l$).
%
\subsection{Safeguarded binary partition}
\label{ss:sbs}
Our first modification of the classic scheme of Sedgewick
\cite[\S5.2.2, Algorithm Q]{knu:acpIII2} proceeds as follows.
After comparing the pivot $v:=x_l$ to $x_r$ to produce the initial
setup
\begin{equation}
\begin{tabular}{llrlrlrr}
\hline
\multicolumn{1}{|c|}{$x=v$} &
\multicolumn{2}{|c|}{$x<v$} &
\multicolumn{2}{|c|}{?} &
\multicolumn{2}{|c|}{$x>v$} &
\multicolumn{1}{|c|}{$x=v$} \\
\hline
\vphantom{$1^{{2^3}^4}$} 
$l$ & $p$ & $i$ & & & $j$ & $q$ & $r$\\
\end{tabular}
\label{ternini}
\end{equation}
with $i:=l$ and $j:=r$,
we work with the three inner blocks of the array
\begin{equation}
\begin{tabular}{lllrrr}
\hline
\multicolumn{1}{|c|}{$x=v$} &
\multicolumn{1}{|c|}{$x\le v$} &
\multicolumn{2}{|c|}{?} &
\multicolumn{1}{|c|}{$x\ge v$} &
\multicolumn{1}{|c|}{$x=v$}\\
\hline
\vphantom{$1^{{2^3}^4}$} 
$l$ & $p$ & $i$ & $j$ & $q$ & $r$\\
\end{tabular}
\label{binbeg}
\end{equation}
until the middle part is empty or just contains an element equal to the
pivot
\begin{equation}
\begin{tabular}{llrclrr}
\hline
\multicolumn{1}{|c|}{$x=v$} &
\multicolumn{2}{|c|}{$x\le v$} &
\multicolumn{1}{|c|}{$x=v$} &
\multicolumn{2}{|c|}{$x\ge v$} &
\multicolumn{1}{|c|}{$x=v$} \\
\hline
\vphantom{$1^{{2^3}^4}$} 
$l$ & $p$ & $j$ & & $i$ & $q$ & $r$ \\
\end{tabular}
\label{binmid}
\end{equation}
(i.e., $j=i-1$ or $j=i-2$),
then swap the ends into the middle for the final arrangement
\begin{equation}
\begin{tabular}{llrr}
\hline
\multicolumn{1}{|c|}{$x\le v$} &
\multicolumn{2}{|c|}{$x=v$} &
\multicolumn{1}{|c|}{$x\ge v$}\\
\hline
\vphantom{$1^{{2^3}^4}$} 
$l$ & $a$ & $b$ & $r$\\
\end{tabular}\ .
\label{binend}
\end{equation}
%
\begin{scheme}[Safeguarded binary partition]
\label{sbs}
\rm
\begin{description}
\itemsep0pt
\item[]
\item[\ref{sbs}1.] [Initialize.]
Set $i:=l$, $p:=i+1$, $j:=r$ and $q:=j-1$.
If $v>x_j$, exchange $x_i\leftrightarrow x_j$ and set $p:=i$;
else if $v<x_j$, set $q:=j$.
\item[\ref{sbs}2.] [Increase $i$ until $x_i\ge v$.]
Increase $i$ by $1$; then if $x_i<v$, repeat this step.
\item[\ref{sbs}3.] [Decrease $j$ until $x_j\le v$.]
Decrease $j$ by $1$; then if $x_j>v$, repeat this step.
\item[\ref{sbs}4.] [Exchange.]
(Here $x_j\le v\le x_i$.)
If $i<j$, exchange $x_i\leftrightarrow x_j$ and return to \ref{sbs}2.
If $i=j$ (so that $x_i=x_j=v$), increase $i$ by $1$ and decrease $j$
by $1$.
\item[\ref{sbs}5.] [Cleanup.]
Set $a:=l+j-p+1$ and $b:=r-q+i-1$.  If $l<p$, exchange
$x_l\leftrightarrow x_j$.  If $q<r$, exchange $x_i\leftrightarrow x_r$.
\end{description}
\end{scheme}

Step \ref{sbs}1 ensures that $x_i\le v\le x_j$, so steps \ref{sbs}2
and \ref{sbs}3 don't need to test whether $i\le j$.  In other words,
while searching for a pair of elements to exchange, the previously
sorted data (initially, $x_l\le x_r$) are used to bound the search,
and the index values are compared only when an exchange is to be
made.  This leads to a small amount of overshoot in the search: in
addition to the necessary $n-1$ comparisons, scheme \ref{sbs} makes
two spurious comparisons or just one (when $i=j+1$ or $i=j$ at
\ref{sbs}4 respectively).  Step \ref{sbs}4 makes at
most $n/2$ index comparisons and at most $n/2-1$ swaps (since
$j-i$ decreases at least by $2$ between swaps); thus \ref{sbs}1 and
\ref{sbs}4 make at most $n/2$ swaps.  To avoid vacuous swaps, step
\ref{sbs}5 may use the tests $l<\min\{p,j\}$ and $\max\{q,i\}<r$;
on the other hand, \ref{sbs}5 could make unconditional swaps
without impairing \eqref{binend}.

Of course, scheme \ref{sbs} could be described in other equivalent
ways.  For instance, \ref{sbs}1 and \ref{sbs}5 can be written in
terms of binary variables $i_l:=p-l$ and $i_r:=r-q$; then
\ref{sbs}5 may decrease $j$ by $1$ if $i_l=1$ and increase $i$ by
$1$ if $i_r=1$ to have $a=j+1$, $b=i-1$ in \eqref{binend}.

A more drastic simplification could swap $x_l\leftrightarrow x_r$
if $v>x_r$ at \ref{sbs}1, omit the second instruction of \ref{sbs}4,
set $a:=b:=j$ at \ref{sbs}5 and swap $x_l\leftrightarrow x_j$ if
$x_l=v$, $x_j\leftrightarrow x_r$ otherwise.
%
\subsection{Single-index controlled binary partition}
\label{ss:sbind1}
It is instructive to compare scheme \ref{sbs} with a popular
scheme of Sedgewick \cite[Progs.\ 3 and 4]{bemc:esf}, based on the
arrangements \eqref{binbeg}--\eqref{binmid} with $p:=l+1$, $q:=r$.
%
\begin{scheme}[Single-index controlled binary partition]
\label{sbind1}
\rm
\begin{description}
\itemsep0pt
\item[]
\item[\ref{sbind1}1.] [Initialize.]
Set $i:=l$ and $j:=r+1$.
\item[\ref{sbind1}2.] [Increase $i$ until $x_i\ge v$.]
Increase $i$ by $1$; then if $i\le r$ and $x_i<v$, repeat this step.
\item[\ref{sbind1}3.] [Decrease $j$ until $x_j\le v$.]
Decrease $j$ by $1$; then if $x_j>v$, repeat this step.
\item[\ref{sbind1}4.] [Exchange.]
(Here $x_j\le v\le x_i$.)
If $i\le j$, exchange $x_i\leftrightarrow x_j$ and return to
\ref{sbind1}2.
\item[\ref{sbind1}5.] [Cleanup.]
Exchange $x_l\leftrightarrow x_j$.
\end{description}
\end{scheme}

The test $i\le r$ of step \ref{sbind1}2 is necessary when $v$ is
greater than the remaining elements.  If $i=j$ at \ref{sbind1}4, a
vacuous swap is followed by one or two unnecessary comparisons; hence
\ref{sbind1}4 may be replaced by \ref{sbs}4 to achieve the same effect
at no extra cost.  With this replacement, scheme \ref{sbind1} makes
$n+1$ comparisons or $n$ if $i=j$ or $i=r+1$ at \ref{sbind1}4, and at
most $(n+1)/2$ index comparisons and $(n-1)/2$ swaps at \ref{sbind1}4.
Usually scheme \ref{sbind1} is used as if $a:=b:=j$ in \eqref{binend},
but in fact \ref{sbind1}5 may set $a:=j$, $b:=i-1$ (note that the final
arrangement of \cite[p.\ 1252]{bemc:esf} is wrong when $j=i-2$).
Therefore, from now on, we assume that scheme \ref{sbind1} incorporates
our suggested modifications of steps \ref{sbind1}4 and \ref{sbind1}5.
%
\subsection{Double-index controlled binary partition}
\label{ss:sbind2}
The following scheme compares both scanning indices $i$ and $j$ in
their inner loops.
%
\begin{scheme}[Double-index controlled binary partition]
\label{sbind2}
\rm
\begin{description}
\itemsep0pt
\item[]
\item[\ref{sbind2}1.] [Initialize.]
Set $i:=l+1$ and $j:=r$.
\item[\ref{sbind2}2.] [Increase $i$ until $x_i\ge v$.]
If $i\le j$ and $x_i<v$, increase $i$ by $1$ and repeat this step.
\item[\ref{sbind2}3.] [Decrease $j$ until $x_j\le v$.]
If $i<j$ and $x_j>v$, decrease $j$ by $1$ and repeat this step.
If $i\ge j$, set $j:=i-1$ and go to \ref{sbind2}5.
\item[\ref{sbind2}4.] [Exchange.]
Exchange $x_i\leftrightarrow x_j$, increase $i$ by $1$, decrease
$j$ by $1$ and return to \ref{sbind2}2.
\item[\ref{sbind2}5.] [Cleanup.]
Set $a:=b:=j$.  Exchange $x_l\leftrightarrow x_j$.
\end{description}
\end{scheme}

Thanks to its tight index control, scheme \ref{sbind2} makes just
$n-1$ comparisons and at most $(n-1)/2$ swaps at \ref{sbind2}4.
Suprisingly, we have not found this scheme in the literature.
%
\subsection{Lomuto's binary partition}
\label{ss:sbL}
We now consider Lomuto's partition
\cite[Prog.\ 2]{bemc:esf}, based on the arrangements
\begin{equation}
\begin{tabular}{crclr}
\hline
\multicolumn{1}{|c|}{$v$} &
\multicolumn{1}{|c|}{$x<v$} &
\multicolumn{1}{|c|}{$x\ge v$} &
\multicolumn{2}{|c|}{$?$} \\
\hline
\vphantom{$1^{{2^3}^4}$} 
$l$ & $p$ & & $i$ & $r$ \\
\end{tabular}
\quad\longrightarrow\quad
\begin{tabular}{crr}
\hline
\multicolumn{1}{|c|}{$v$} &
\multicolumn{1}{|c|}{$x<v$} &
\multicolumn{1}{|c|}{$x\ge v$} \\
\hline
\vphantom{$1^{{2^3}^4}$} 
$l$ & $p$ & $r$ \\
\end{tabular}\ .
\label{sbLmid}
\end{equation}
%
\begin{scheme}[Lomuto's binary partition]
\label{sbL}
\rm
\begin{description}
\itemsep0pt
\item[]
\item[\ref{sbL}1.] [Initialize.]
Set $i:=l+1$ and $p:=l$.
\item[\ref{sbL}2.] [Check if done.]
If $i>r$, go to \ref{sbL}4.
\item[\ref{sbL}3.] [Exchange if necessary.]
If $x_i<v$, increase $p$ by $1$ and exchange $x_p\leftrightarrow x_i$.
Increase $i$ by $1$ and return to \ref{sbL}2.
\item[\ref{sbL}4.] [Cleanup.]
Set $a:=b:=p$.  Exchange $x_l\leftrightarrow x_p$.
\end{description}
\end{scheme}

At the first sight, scheme \ref{sbL} looks good: it makes just the
$n-1$ necessary comparisons.  However, it can make up to $n-1$ swaps
(e.g., vacuous swaps when $v$ is greater than the remaining elements,
or $n-2$ nonvacuous swaps for $x[l\colon r]=[n-1,n,1,2,\ldots,n-2]$).
%
\subsection{Comparison of bipartitioning schemes}
\label{ss:compbi}
%
\subsubsection{Swaps for distinct keys}
\label{ss:compbidist}
When the elements are distinct, we have strict inequalities in
\eqref{binbeg}--\eqref{sbLmid}, $j=i-1$ in \eqref{binmid} and
$a=b$ in \eqref{binend}.  Distinguishing {\em low\/} keys
$x_m<v$ and {\em high\/} keys $x_m>v$, let $t$ be the number of
high keys in the input subarray $x[l+1\colon a]$.  Then schemes
\ref{sbind1} and \ref{sbind2} make the {\em same\/} sequence of $t$
swaps to produce the arrangement
\begin{equation}
\begin{tabular}{crr}
\hline
\multicolumn{1}{|c|}{$v$} &
\multicolumn{1}{|c|}{$x<v$} &
\multicolumn{1}{|c|}{$x>v$}\\
\hline
\vphantom{$1^{{2^3}^4}$} 
$l$ & $a$ & $r$\\
\end{tabular}
\label{sbind1ind2}
\end{equation}
before the final swap $x_l\leftrightarrow x_a$, and their operation
is described by the instruction: until there are no high keys in
$x[l+1\colon a]$, swap the leftmost high key in $x[l+1\colon a]$ with
the rightmost low key in $x[a+1\colon r]$.  Thus schemes
\ref{sbind1} and \ref{sbind2} make just the necessary $t$ swaps.
Scheme \ref{sbs} acts in the same way if $x_r>v$ at \ref{sbs}1.
If $x_r<v$ at \ref{sbs}1, let $t_l$ be the number of low keys in
$x[a\colon r]$; in this {\em low\/} case, after the initial swap
$x_l\leftrightarrow x_r$, scheme \ref{sbs} makes $t_l-1$ swaps,
each time exchanging the leftmost high key in $x[l+1\colon a-1]$ with
the rightmost low key in $x[a\colon r-1]$, to produce the arrangement
\begin{equation}
\begin{tabular}{llc}
\hline
\multicolumn{1}{|c|}{$x<v$} &
\multicolumn{1}{|c|}{$x>v$} &
\multicolumn{1}{|c|}{$v$} \\
\hline
\vphantom{$1^{{2^3}^4}$} 
$l$ & $a$ & $r$\\
\end{tabular}
\label{sbslow}
\end{equation}
before the final swap $x_a\leftrightarrow x_r$.  Since the number of
low keys in $x[a+1\colon r]$ equals $t$, we have $t_l=t+1$ if $x_a<v$,
otherwise $t_l=t$.  Thus, relative to schemes \ref{sbind1} and
\ref{sbind2}, scheme \ref{sbs} makes an extra swap when both $x_a$
and $x_r$ are low.  Note that schemes \ref{sbs}, \ref{sbind1} and
\ref{sbind2} never swap the same key twice while producing the
arrangements \eqref{sbind1ind2}--\eqref{sbslow}.  In contrast, scheme
\ref{sbL} may swap the same high key many times while producing the
arrangement \eqref{sbind1ind2} (usually different from that of
\ref{sbind1} and \ref{sbind2}).  In fact scheme \ref{sbL} makes
exactly $t_{\rm\ref{sbL}}:=a-l$ swaps; this is the total number of
low keys.  Thus the number of extra swaps made by scheme \ref{sbL}
relative to \ref{sbind1} and \ref{sbind2}, $t_{\rm\ref{sbL}}-t$,
equals the number of low keys in the initial $x[l+1\colon a]$.
%
\subsubsection{Swaps for equal keys}
\label{ss:compbieq}
When equal keys occur, schemes \ref{sbs}, \ref{sbind1} and \ref{sbind2}
perform similarly to Sedgewick's scheme of \cite[Prog.\ 1]{sed:qek};
in particular, thanks to stopping the scanning pointers on keys equal
to the pivot, they tend to produce balanced partitions.  For instance,
when all the keys are equal, we get the following partitions:
for scheme \ref{sbs}, $a=\lfloor(l+r-1)/2\rfloor$, $b=a+1+(n\bmod 2)$
after $\lceil(n+1)/2\rceil$ swaps; for scheme \ref{sbind1},
$a=\lceil(l+r)/2\rceil$, $b=a+1-(n\bmod 2)$ after $\lceil n/2\rceil$
swaps; for scheme \ref{sbind2}, $a=b=\lceil(l+r)/2\rceil$ after
$\lceil n/2\rceil$ swaps.
In contrast, scheme \ref{sbL} makes no swaps, but yields $a=b=l$, the
worst possible partition.
%
\section{Average-case analysis of bipartitioning schemes}
\label{s:averbi}
In this section we assume that the keys to be partitioned are distinct
and in random order; since the schemes depend only on the relative
order of the keys, we may as well assume that they are the first $n$
positive integers in random order.
For simplier notation, we suppose that $l=1$ and $r=n$.
It is easy to see that when the keys in $x[l+1\colon r]$ are
in random order, each scheme of \S\ref{s:binpart} {\em preserves
randomness\/} in the sense of producing $x[l\colon a-1]$ and
$x[a+1\colon r]$ in which the low and high keys are in random order
(since the relative orders of the low keys and the high keys on input
have no effect on the scheme).
%
\subsection{Expected numbers of swaps for fixed pivot ranks}
\label{ss:expswapfixed}
For a given pivot $v:=x_1$, let $j_v$ denote the number of low keys
in the array $x[2\colon n]$; then $a=j_v+1$ is the rank of $v$.  Once
$j_v$ is fixed at $j$ (say), to compute the average number of swaps
made by each scheme, it's enough to assume that the keys in
$x[2\colon n]$ are in random order; thus averages are taken over the
$(n-1)!$ distinct inputs.  Our analysis hinges on the following
well-known fact (cf.\ \cite{chv:aaq}).
%
\begin{fact}
\label{f:randlow}
Suppose an array\/ $x[\hat l\colon\hat r]$ contains\/
$\hat n:=\hat r-\hat l+1>0$ distinct keys, of which\/ $\hat\jmath$
are low and\/ $\hat n-\hat\jmath$ are high.  If all the\/ $\hat n!$
permutations of the keys are equiprobable, then\/
$\hat\jmath(\hat n-\hat\jmath)/\hat n$ is the average number of high
keys in the first\/ $\hat\jmath$ positions.
\end{fact}
\begin{proof}
List all the $\hat n!$ key permutations as rows of an
$\hat n!\times\hat n$ matrix.  In each column, each key appears
$(\hat n-1)!$ times, so the number of high keys in the first
$\hat\jmath$ columns is $\hat\jmath(\hat n-\hat\jmath)(\hat n-1)!$;
dividing by $\hat n!$ gives the average number
$\hat\jmath(\hat n-\hat\jmath)/\hat n$.
\qed
\end{proof}
%
\begin{lemma}
\label{l:expswapj}
Suppose the number of low key equals\/ $j$.
Let\/ $T_j^{\rm\ref{sbs}}$, $T_j^{\rm\ref{sbind1}}$,
$T_j^{\rm\ref{sbind2}}$, $T_j^{\rm\ref{sbL}}$ denote
the average numbers of swaps made by schemes\/
{\rm\ref{sbs}}, {\rm\ref{sbind1}}, {\rm\ref{sbind2}} and\/
{\rm\ref{sbL}}, excluding the final swaps.  Then
\begin{subequations}
\label{Tjsbs}
\begin{equation}
T_j^{\rm\ref{sbs}}=\frac{j(n-1-j)}{n-1}\frac{n-3}{n-2}+
\frac{j}{n-1},
\quad n\ge 3,
\label{Tjsbs3}
\end{equation}
\begin{equation}
T_j^{\rm\ref{sbs}}=\frac{j}{n-1},
\quad n=2,
\label{Tjsbs2}
\end{equation}
\end{subequations}
\begin{equation}
T_j^{\rm\ref{sbind1}}=T_j^{\rm\ref{sbind2}}=\frac{j(n-1-j)}{n-1},
\label{Tjsbind1}
\end{equation}
\begin{equation}
T_j^{\rm\ref{sbL}}=j.
\label{TjsbL}
\end{equation}
\end{lemma}
\begin{proof}
By assumption, the arrangements \eqref{sbind1ind2}--\eqref{sbslow}
involve $l=1$, $a=j+1$, $r=n$.  The results follow from suitable
choices of $\hat l$, $\hat r$, $\hat\jmath$ in Fact \ref{f:randlow}.

For scheme \ref{sbs}, assuming $n\ge3$, let $\hat l=2$, $\hat r=n-1$.
Depending on whether $x_n>v$ or $x_n<v$, scheme \ref{sbs} produces
either \eqref{sbind1ind2} or \eqref{sbslow} from the initial
configurations
\begin{equation}
\begin{tabular}{clrlcrc}
\hline
\multicolumn{1}{|c|}{$v$} &
\multicolumn{2}{|c|}{} &
\multicolumn{3}{|c|}{} &
\multicolumn{1}{|c|}{$x>v$}\\
\hline
\vphantom{$1^{{2^3}^4}$} 
$1$ & & $a$ & & & & $n$\\
\end{tabular}
\qquad\mbox{\rm or}\qquad
\begin{tabular}{clcrlrc}
\hline
\multicolumn{1}{|c|}{$v$} &
\multicolumn{3}{|c|}{} &
\multicolumn{2}{|c|}{} &
\multicolumn{1}{|c|}{$x<v$}\\
\hline
\vphantom{$1^{{2^3}^4}$} 
$1$ & & & & $a$ & & $n$\\
\end{tabular}\ .
\label{sbsx<vx>v}
\end{equation}
For $x_n>v$, take $\hat\jmath=j=a-1$; then the average number of
high keys in $x[2\colon a]$ (i.e., of swaps) equals
$j(n-2-j)/(n-2)$.  For $x_n<v$, take $\hat\jmath=j-1$; in this case
$t_l-1$, the number of low keys in $x[a\colon n-1]$, equals the
number of high keys in $x[2\colon j]$, so the average value of
$t_l$ equals $(j-1)(n-1-j)/(n-2)+1$.  Since there are $j$ low keys
and $n-1-j$ high keys which appear in random order, we have $x_n>v$
with probability $(n-1-j)/(n-1)$ and $x_n<v$ with probability
$j/(n-1)$.  Adding the contributions of these cases multiplied
by their probabilities yields \eqref{Tjsbs3}.  For $n=2$, \ref{sbs}1
makes $1$ swap if $j=1$, $0$ otherwise, so \eqref{Tjsbs2} holds.

For schemes \ref{sbind1} and \ref{sbind2}, take $\hat l=2$, $\hat r=n$,
$\hat\jmath=j$ to get \eqref{Tjsbind1} in a similar way.

Since scheme \ref{sbL} makes $t_{\rm\ref{sbL}}:=a-l=j$ swaps,
\eqref{TjsbL} follows.
\qed
\end{proof}

To compare the average values \eqref{Tjsbs}--\eqref{TjsbL}, note that
we have $0\le j\le n-1$,
\begin{equation}
T_j^{\rm\ref{sbs}}=T_j^{\rm\ref{sbind1}}+\frac{j(j-1)}{(n-1)(n-2)}
\quad\mbox{\rm and}\quad
T_j^{\rm\ref{sbL}}=T_j^{\rm\ref{sbs}}+
\frac{j+(n-3)j^2}{(n-1)(n-2)}
\quad\mbox{\rm if}\ n\ge3,
\label{Tjsbcomp}
\end{equation}
$T_j^{\rm\ref{sbind1}}=0$ and $T_j^{\rm\ref{sbs}}=T_j^{\rm\ref{sbL}}=j$
if $n=2$.  Thus $T_j^{\rm\ref{sbind1}}\le T_j^{\rm\ref{sbs}}+1$ (with
equality iff there are no high keys), whereas $T_j^{\rm\ref{sbL}}$
is much greater than $T_j^{\rm\ref{sbs}}$ when there are relatively
many low keys.
%
\subsection{Bounding expected numbers of swaps for arbitrary pivots}
\label{ss:expswapany}
From now on we assume that the pivot is selected by an arbitrary rule
for which (once the pivot is swapped into $x_l$ if necessary) each
permutation of the remaining keys is equiprobable.
Let $T_{\rm\ref{sbs}}$, $T_{\rm\ref{sbind1}}$, $T_{\rm\ref{sbind2}}$,
$T_{\rm\ref{sbL}}$ denote the average numbers of swaps made by schemes
\ref{sbs}, \ref{sbind1}, \ref{sbind2} and \ref{sbL}, {\em excluding\/}
the final swaps.  Of course, these numbers depend on details of pivot
selection, but they can be bounded independently of such details.
To this end we compute the maxima of the average values
\eqref{Tjsbs}--\eqref{TjsbL}.
%
\begin{lemma}
\label{l:expswapmax}
Let\/ $T_{\rm max}^{\rm\ref{sbs}}$, $T_{\rm max}^{\rm\ref{sbind1}}$,
$T_{\rm max}^{\rm\ref{sbind2}}$, $T_{\rm max}^{\rm\ref{sbL}}$ denote
the maxima of\/ $T_j^{\rm\ref{sbs}}$, $T_j^{\rm\ref{sbind1}}$,
$T_j^{\rm\ref{sbind2}}$, $T_j^{\rm\ref{sbL}}$ over\/ $0\le j<n$.
Then
\begin{equation}
\renewcommand{\arraystretch}{1.5}
T_{\rm max}^{\rm\ref{sbs}}=
\left\{\begin{array}{ll}
\displaystyle
\frac{n}4-\frac{(n-5)(n\bmod 2)}{4(n-1)(n-2)}&\mbox{if\/ $n\ge5$},\\
\;1&\mbox{if\/ $n\le 4$},
\end{array}\right.
\label{Tmaxsbs}
\end{equation}
\begin{equation}
T_{\rm max}^{\rm\ref{sbind1}}=T_{\rm max}^{\rm\ref{sbind2}}=
\frac{n-1}4-\frac{(n+1)\bmod 2}{4(n-1)},
\label{Tmaxsbind1}
\end{equation}
\begin{equation}
T_{\rm max}^{\rm\ref{sbL}}=n-1.
\label{TmaxsbL}
\end{equation}
\end{lemma}
\begin{proof}
The maximum of \eqref{Tjsbs} is attained at $j=\lceil n/2\rceil$ if
$n\ge4$, $j=n-1$ otherwise.
The maximum of \eqref{Tjsbind1} is attained at $j=\lfloor n/2\rfloor$.
The rest follows by simple computations.
\qed
\end{proof}
%
\begin{corollary}
\label{c:expswapmax}
The average numbers of swaps\/
$T_{\rm\ref{sbs}}$, $T_{\rm\ref{sbind1}}$,
$T_{\rm\ref{sbind2}}$, $T_{\rm\ref{sbL}}$ made by schemes\/
{\rm\ref{sbs}}, {\rm\ref{sbind1}}, {\rm\ref{sbind2}}, {\rm\ref{sbL}}
are at most\/ $T_{\rm max}^{\rm\ref{sbs}}$,
$T_{\rm max}^{\rm\ref{sbind1}}$, $T_{\rm max}^{\rm\ref{sbind2}}$,
$T_{\rm max}^{\rm\ref{sbL}}$ for the values given in\/
\eqref{Tmaxsbs}--\eqref{TmaxsbL}.  In particular,
$T_{\rm\ref{sbs}}$, $T_{\rm\ref{sbind1}}$ and\/
$T_{\rm\ref{sbind2}}$ are at most\/ $n/4$ for\/ $n>3$.
\end{corollary}
%
\subsection{The case where pivots are chosen via sampling}
\label{ss:pivsample}
%
\subsubsection{Pivots with fixed sample ranks}
\label{ss:pivsamplefixed}
We assume that the pivot $v$ is selected as the $(p+1)$th element in
a sample of size $s$, $0\le p<s\le n$.  Thus $p$ and
$q:=s-1-p$ are the numbers of low and high keys in the sample,
respectively.  Recall that $v$ has rank $j_v+1$, where $j_v$ is the
total number of low keys.  We shall need the following two expected
values for this selection:
\begin{equation}
\Exp j_v=E(n,s,p):=(p+1)(n+1)/(s+1)-1,
\label{Expjv}
\end{equation}
\begin{equation}
\Exp\left[\frac{j_v(n-1-j_v)}{n-1}\right]=
T(n,s,p):=
\frac{(p+1)(q+1)}{(s+1)(s+2)}\frac{(n+1)(n+2)}{n-1}-
\frac{n}{n-1}.
\label{Expjvn-1jv}
\end{equation}
Here \eqref{Expjv} follows from \cite[Eq.\ (10)]{flri:etb} and
\eqref{Expjvn-1jv} from the proof of \cite[Lem.\ 1]{maro:oss}.
%
\begin{theorem}
\label{t:expswap}
For $E(n,s,p)$ and\/ $T(n,s,p)$ given by\/
\eqref{Expjv}--\eqref{Expjvn-1jv}, the average numbers of swaps\/
$T_{\rm\ref{sbs}}$, $T_{\rm\ref{sbind1}}$,
$T_{\rm\ref{sbind2}}$, $T_{\rm\ref{sbL}}$
made by schemes\/ {\rm\ref{sbs}}, {\rm\ref{sbind1}}, {\rm\ref{sbind2}},
{\rm\ref{sbL}} are equal to, respectively,
\begin{equation}
T_{\rm\ref{sbs}}(n,s,p)=\frac{\max\{n-3,0\}}{\max\{n-2,1\}}T(n,s,p)+
\frac{1}{n-1}E(n,s,p),
\label{Tsbs}
\end{equation}
\begin{equation}
T_{\rm\ref{sbind1}}(n,s,p)=T_{\rm\ref{sbind2}}(n,s,p)=T(n,s,p),
\label{Tsbind1}
\end{equation}
\begin{equation}
T_{\rm\ref{sbL}}(n,s,p)=E(n,s,p).
\label{TsbL}
\end{equation}
\end{theorem}
\begin{proof}
Take expectations of the averages \eqref{Tjsbs}--\eqref{TjsbL}
conditioned on $j_v=j$, and use \eqref{Expjv}--\eqref{Expjvn-1jv};
the two ``$\max$'' operations in \eqref{Tsbs} combine the cases of
$n=2$ and $n\ge3$.
\qed
\end{proof}

The average values \eqref{Tsbs}--\eqref{TsbL} may be compared as
follows.  First, in the classic case of $s=1$ ($p=q=0$), we have
$T_{\rm\ref{sbs}}=n/6$ if $n\ge3$ (else $T_{\rm\ref{sbs}}=1/2$),
$T_{\rm\ref{sbind1}}=(n-2)/6$, $T_{\rm\ref{sbL}}=(n-1)/2$; thus
scheme \ref{sbL} makes about three times as many swaps as \ref{sbs},
\ref{sbind1} and \ref{sbind2}.

Second, for nontrivial samples ($s>1$) one may ask which choices of
$p$ are ``good'' or ``bad'' with respect to swaps.  For schemes
\ref{sbind1} and \ref{sbind2},
the worst case occurs if $p$ is chosen to maximize \eqref{Expjvn-1jv}
(where $q+1=s-p$); we obtain that for all $0\le p<s$,
\begin{equation}
T(n,s,p)\le\kappa(s)\frac{(n+1)(n+2)}{n-1}-\frac{n}{n-1}\le
\frac{n-1}4
\quad\mbox{\rm with}\quad
\kappa(s):=\frac{s+1}{4(s+2)},
\label{Tkappa}
\end{equation}
where the first inequality holds as equality only for the
{\em median-of-$s$\/} choice of $p=(s-1)/2$, and the second one iff
$s=n$.  Since $T_{\rm\ref{sbs}}\le T_{\rm\ref{sbind1}}+1$,
\eqref{Tkappa} yields $T_{\rm\ref{sbs}}\le(n+3)/4$, but we already
know that $T_{\rm\ref{sbs}}\le n/4$ (Cor.\ \ref{c:expswapmax}).
For any median-of-$s$ choice with a fixed $s$, $T_{\rm\ref{sbs}}$ and
$T_{\rm\ref{sbind1}}$
are asymptotically $\kappa(s)n$, whereas $E(n,s,p)=(n-1)/2$; thus
scheme \ref{sbL} makes about $1/2\kappa(s)>2$ times as many swaps as
\ref{sbs}, \ref{sbind1} and \ref{sbind2} (with $\kappa(3)=1/5$,
$\kappa(5)=3/14$, $\kappa(7)=2/9$, $\kappa(9)=5/22$).  On the other
hand, for the extreme choices of $p=0$ or $p=s-1$ which minimize
\eqref{Expjvn-1jv} (then $v$ is the smallest or largest key in
the sample), $T_{\rm\ref{sbs}}$ and $T_{\rm\ref{sbind1}}$ are
asymptotically $ns/(s+1)(s+2)$, whereas $T_{\rm\ref{sbL}}$ is
asymptotically $n/(s+1)$ for $p=0$ and $ns/(s+1)$ for $p=s-1$.
Thus scheme \ref{sbL} can't improve upon \ref{sbs} and \ref{sbind1}
even for the choice of $p=0$ which minimizes \eqref{Expjv}.
%
\subsubsection{Pivots with random sample ranks}
\label{ss:pivsamplerand}
Following the general framework of \cite[\S1]{chhwts:atc}, suppose the
pivot $v$ is selected by taking a random sample of $s$ elements,
and choosing the $(p+1)$th element in this sample with probability
$\pi_p$, $0\le p<s$, $\sum_{p=0}^{s-1}\pi_p=1$.  In other words, for
$p_v$ denoting the number of low keys in the sample, we have
$\Pr[p_v=p]=\pi_p$.  Hence, by viewing \eqref{Expjv}--\eqref{TsbL}
as expectations conditioned on the event $p_v=p$, we may take total
averages to get
\begin{equation}
\Exp j_v=\Exp\left[E(n,s,p_v)\right]=E(n,s):=
(\Exp p_v+1)(n+1)/(s+1)-1,
\label{Expjvp}
\end{equation}
\begin{equation}
\Exp\left[\frac{j_v(n-1-j_v)}{n-1}\right]=
\Exp\left[T(n,s,p_v)\right]=T(n,s):=
\sum_{0\le p<s}\pi_pT(n,s,p),
\label{Expjvn-1jvp}
\end{equation}
and the following extension of Theorem \ref{t:expswap}.
%
\begin{theorem}
\label{t:expswapp}
For $E(n,s)$ and\/ $T(n,s)$ given by\/
\eqref{Expjvp}--\eqref{Expjvn-1jvp}, the average numbers of swaps\/
$T_{\rm\ref{sbs}}$, $T_{\rm\ref{sbind1}}$,
$T_{\rm\ref{sbind2}}$, $T_{\rm\ref{sbL}}$
made by schemes\/ {\rm\ref{sbs}}, {\rm\ref{sbind1}}, {\rm\ref{sbind2}},
{\rm\ref{sbL}} are equal to, respectively,
\begin{equation}
T_{\rm\ref{sbs}}(n,s)=\frac{\max\{n-3,0\}}{\max\{n-2,1\}}T(n,s)+
\frac{1}{n-1}E(n,s),
\label{Tsbsp}
\end{equation}
\begin{equation}
T_{\rm\ref{sbind1}}(n,s)=T_{\rm\ref{sbind2}}(n,s)=T(n,s),
\label{Tsbind1p}
\end{equation}
\begin{equation}
T_{\rm\ref{sbL}}(n,s)=E(n,s).
\label{TsbLp}
\end{equation}
\end{theorem}

Note that in \eqref{Expjvp}--\eqref{Expjvn-1jvp}, we have
$\Exp p_v=\sum_{0\le p<s}\pi_pp\le s-1$ and
\begin{equation}
T(n,s)=\check\kappa(s)\frac{(n+1)(n+2)}{n-1}-\frac{n}{n-1}
\quad\mbox{\rm with}\quad
\check\kappa(s):=\sum_{0\le p<s}\pi_p\frac{(p+1)(s-p)}{(s+1)(s+2)},
\label{Tkappaavg}
\end{equation}
where $\check\kappa(s)\le\kappa(s)$ (cf.\ \eqref{Tkappa}), and
$\check\kappa(s)=\kappa(s)$ iff $\pi_p=1$ for $p=(s-1)/2$.  Thus
again $T_{\rm\ref{sbs}}$ and $T_{\rm\ref{sbind1}}$ are asymptotically
$\check\kappa(s)n$, whereas $T_{\rm\ref{sbL}}$ can be much larger.

As an important example, we consider {\em Tukey's ninther\/}, the
median of three elements each of which is the median of three elements
\cite{bemc:esf}.  Then $s=9$ and $\pi_p=0$ except for
$\pi_3=\pi_5=3/14$, $\pi_4=3/7$ \cite{chhwts:atc,dur:aao}, so
$E(n,9)=(n-1)/2$ and $\check\kappa(9)=86/385\approx0.223$.  Thus,
when the ninther replaces the median-of-3, $T_{\rm\ref{sbs}}$ and
$T_{\rm\ref{sbind1}}$ increase by about $12$ percent, getting closer
to $n/4$, whereas $T_{\rm\ref{sbL}}$ stays at $(n-1)/2$.
%
\section{Using sample elements as sentinels}
\label{s:sampsent}
The schemes of \S\ref{s:binpart} can be tuned \cite[\S7.2]{maro:oss}
when the pivot $v$ is selected as the $(p+1)$th element in a sample of
size $s$, assuming $0\le p<s\le n$ and $q:=s-1-p>0$.

First, suppose the $p$ sample keys $\le v$ are placed first, followed
by $v$, and the remaining $q$ sample keys $\ge v$ are placed at
the end of the array $x[l\colon r]$.  Then, for $\bar l:=l+p$ and
$\bar r:=r-q$, we only need to partition the array
$x[\bar l\colon\bar r]$ of size $\bar n:=n-s+1$.  The schemes of
\S\ref{s:binpart} are modified as follows.

In step \ref{sbs}1 of scheme \ref{sbs}, set $i:=\bar l$ and
$j:=\bar r+1$; in step \ref{sbs}5 set $a:=j$, $b:=i-1$ and exchange
$x_{\bar l}\leftrightarrow x_j$.  This scheme makes $\bar n+1$
comparisons, or just $\bar n$ if $i=j$ at \ref{sbs}4.  The same scheme
results from scheme \ref{sbind1} by replacing $l$, $r$ with $\bar l$,
$\bar r$, \ref{sbind1}4 with \ref{sbs}4, and omitting the test
``$i\le r$'' in \ref{sbind1}2.  Similarly, $\bar l$, $\bar r$ replace
$l$ and $r$ in schemes \ref{sbind2} and \ref{sbL}, which make
$\bar n-1$ comparisons.

To extend the results of \S\ref{s:averbi} to these modifications, note
that for $\bar n=1$ these schemes make no swaps except for the final
ones.  For $\bar n>1$, schemes \ref{sbs}, \ref{sbind1} and \ref{sbind2}
swap the same keys, if any.  Therefore, under the sole assumption that
the keys in $x[\bar l+1\colon\bar r]$ are distinct and in random order,
Lemma \ref{l:expswapj} holds with \eqref{Tjsbs}--\eqref{TjsbL} replaced
by
\begin{equation}
T_j^{\rm\ref{sbs}}=T_j^{\rm\ref{sbind1}}=T_j^{\rm\ref{sbind2}}=
\frac{(j-p)(n-1-q-j)}{n-s}
\qquad\mbox{\rm and}\qquad
T_j^{\rm\ref{sbL}}=j-p,
\label{TjsbssbL}
\end{equation}
using $\hat l=\bar l+1$, $\hat r=\bar r$, $\hat\jmath=j-p$ in
Fact \ref{f:randlow}; further, Lemma \ref{l:expswapmax} and
Corollary \ref{c:expswapmax} hold with $n$ replaced by $\bar n$,
\eqref{Tmaxsbs} omitted and
$T_{\rm max}^{\rm\ref{sbs}}=T_{\rm max}^{\rm\ref{sbind1}}$ in
\eqref{Tmaxsbind1}.  Next, \eqref{Expjv}--\eqref{Expjvn-1jv} are
replaced by
\begin{equation}
\Exp j_v-p=\hat E(n,s,p):=(p+1)(n-s)/(s+1),
\label{Expjv-p}
\end{equation}
\begin{equation}
\Exp\left[\frac{(j_v-p)(n-1-q-j_v)}{n-s}\right]=
\hat T(n,s,p):=
\frac{(p+1)(q+1)}{(s+1)(s+2)}(n-s-1),
\quad s<n,
\label{Expjvpn-1q}
\end{equation}
where \eqref{Expjvpn-1q} is obtained similarly to \eqref{Expjvn-1jv}
\cite[\S7.2]{maro:oss}.  In view of
\eqref{TjsbssbL}--\eqref{Expjvpn-1q}, Theorem \ref{t:expswap} holds
with $E(n,s,p)$, $T(n,s,p)$ replaced by $\hat E(n,s,p)$,
$\hat T(n,s,p)$, \eqref{Tsbs} omitted and
$T_{\rm\ref{sbs}}(n,s,p)=T_{\rm\ref{sbind1}}(n,s,p)$ in
\eqref{Tsbind1}.  Finally, \eqref{Tkappa} is replaced by
\begin{equation}
\hat T(n,s,p)\le\kappa(s)(n-s-1)<\frac{n-s-1}4
\quad\mbox{\rm with}\quad
\kappa(s):=\frac{s+1}{4(s+2)},
\label{hatTkappa}
\end{equation}
where the equality holds iff $p=(s-1)/2$, in which case
$\hat E(n,s,p)=(n-s)/2$.

\begin{figure}
\footnotesize
\begin{center}
\setlength{\unitlength}{1.0mm}%
\begin{picture}(161,40)(0,17)
\put(0,41){\begin{picture}(161,10)
\put(80,11){\circle{8}}
\put(80,11){\makebox(0,0){$a:c$}}
\put(65.5,1.5){\line(3,2){11}}
\put(94.5,1.5){\line(-3,2){11}}
\put(70,8){\makebox(0,0){$\le$}}
\put(90,8){\makebox(0,0){$>$}}
\put(61.5,0){\circle{8}}
\put(61.5,0){\makebox(0,0){$a:b$}}
\put(98.5,0){\circle{8}}
\put(98.5,0){\makebox(0,0){$a:b$}}
\end{picture}}
\put(57.5,33.5){\line(2,3){2.45}}
\put(70,32.5){\line(-6,5){6}}
\put(90,32.5){\line(6,5){6}}
\put(102.5,33.5){\line(-2,3){2.45}}
\put(0,27.5){\begin{picture}(152,5)
\put(55.5,2.5){\circle{8}}
\put(55.5,2.5){\makebox(0,0){$b:c$}}
\put(62.5,0){\framebox(16,5){$b<a\le c$}}
\put(81.5,0){\framebox(16,5){$c<a\le b$}}
\put(104.5,2.5){\circle{8}}
\put(104.5,2.5){\makebox(0,0){$b:c$}}
\end{picture}}
\put(0,17.5){\begin{picture}(161,5)
\put(38,0){\framebox(16,5){$a\le b\le c$}}
\put(57,0){\framebox(16,5){$a\le c<b$}}
\put(87,0){\framebox(16,5){$b\le c<a$}}
\put(106,0){\framebox(16,5){$c<b<a$}}
\end{picture}}
\put(46,22.5){\line(6,5){6.0}}
\put(65,22.5){\line(-6,5){6.0}}
\put(95,22.5){\line(6,5){6}}
\put(114,22.5){\line(-6,5){6}}
\end{picture}
\end{center}
\caption{Decision tree for median of three}
\label{fig:med3pm}
\end{figure}
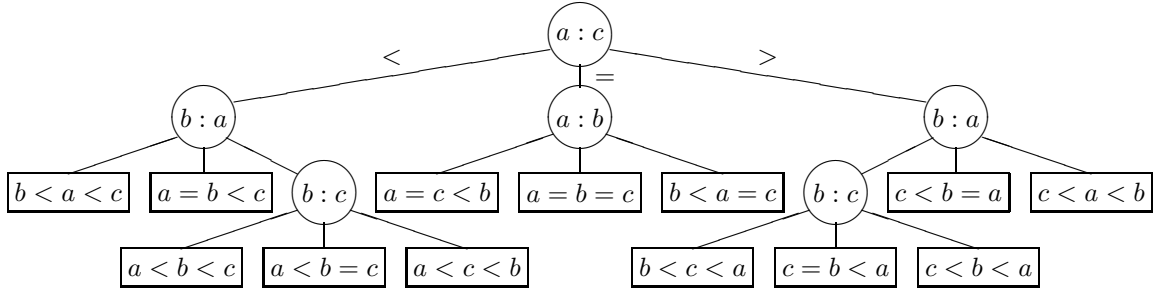

Randomness may be lost when the sample keys are rearranged by pivot
selection, but it is preserved for the median-of-3 selection with
$p=q=1$.  Then the sample keys usually are $x_l$, $x_{l+1}$, $x_r$
(after exchanging $x_{l+1}$ with the middle key
$x_{\lfloor(l+r)/2\rfloor}$).  Arranging the sample according to
Figure \ref{fig:med3pm}
%
takes $8/3$ comparisons and $7/6$ swaps on average for distinct keys.
(These counts hold if, for simpler coding, only the left subtree is
used after exchanging $a\leftrightarrow c$ when $a>c$; other trees
\cite[Prog.\ 5]{bemc:esf} take $3/2$ swaps for such simplifications.)

Even if pivot selection doesn't rearrange the array (except for placing
the pivot in $x_l$), scheme \ref{sbs} may be simplified: in
step \ref{sbs}1, set $i:=l$ and $j:=r+1$; in step \ref{sbs}5 set $a:=j$,
$b:=i-1$ and exchange $x_l\leftrightarrow x_j$.  The same scheme
results from scheme \ref{sbind1} by replacing \ref{sbind1}4 with
\ref{sbs}4, and omitting the test ``$i\le r$'' in \ref{sbind1}2.  This
simplification is justified by the presence of at least one key $\ge v$
in $x[l+1\colon r]$, which stops the scanning index $i$.  Hence the
results of \S\ref{s:averbi} remain valid (with \eqref{Tjsbs},
\eqref{Tjsbcomp}, \eqref{Tmaxsbs}, \eqref{Tsbs}, \eqref{Tsbsp} omitted,
$T_j^{\rm\ref{sbs}}=T_j^{\rm\ref{sbind1}}$ in \eqref{Tjsbind1},
$T_{\rm max}^{\rm\ref{sbs}}=T_{\rm max}^{\rm\ref{sbind1}}$ in
\eqref{Tmaxsbind1},
$T_{\rm\ref{sbs}}(n,s,p)=T_{\rm\ref{sbind1}}(n,s,p)$ in
\eqref{Tsbind1},
$T_{\rm\ref{sbs}}(n,s)=T_{\rm\ref{sbind1}}(n,s)$ in \eqref{Tsbind1p}).
%
\section{Tripartitioning schemes}
\label{s:terpart}
While bipartitioning schemes divide the input keys into $\le v$ and
$\ge v$, tripartitioning schemes divide the keys into $<v$, $=v$ and
$>v$.  We now give ternary versions of the schemes of
\S\ref{s:binpart}, using the following notation for vector swaps
(cf.\ \cite{bemc:esf}).

A vector swap denoted by $x[a\colon b]\leftrightarrow x[b+1\colon c]$
means that the first $d:=\min(b+1-a,c-b)$ elements of array
$x[a\colon c]$ are exchanged with its last $d$ elements in arbitrary
order if $d>0$; e.g., we may exchange
$x_{a+i}\leftrightarrow x_{c-i}$ for $0\le i<d$, or
$x_{a+i}\leftrightarrow x_{c-d+1+i}$ for $0\le i<d$.
%
\subsection{Safeguarded ternary partition}
\label{ss:sbt}
Our ternary version of scheme \ref{sbs} employs the following
``strict'' analogs of \eqref{binbeg}--\eqref{binend}:
\begin{equation}
\begin{tabular}{lllrrr}
\hline
\multicolumn{1}{|c|}{$x=v$} &
\multicolumn{1}{|c|}{$x<v$} &
\multicolumn{2}{|c|}{?} &
\multicolumn{1}{|c|}{$x>v$} &
\multicolumn{1}{|c|}{$x=v$} \\
\hline
\vphantom{$1^{{2^3}^4}$} 
$l$ & $p$ & $i$ & $j$ & $q$ & $r$\\
\end{tabular}\ ,
\label{ternbeg}
\end{equation}
\begin{equation}
\begin{tabular}{llrclrr}
\hline
\multicolumn{1}{|c|}{$x=v$} &
\multicolumn{2}{|c|}{$x<v$} &
\multicolumn{1}{|c|}{$x=v$} &
\multicolumn{2}{|c|}{$x>v$} &
\multicolumn{1}{|c|}{$x=v$} \\
\hline
\vphantom{$1^{{2^3}^4}$} 
$l$ & $p$ & $j$ & & $i$ & $q$ & $r$ \\
\end{tabular}\ ,
\label{ternmid}
\end{equation}
\begin{equation}
\begin{tabular}{llrr}
\hline
\multicolumn{1}{|c|}{$x<v$} &
\multicolumn{2}{|c|}{$x=v$} &
\multicolumn{1}{|c|}{$x>v$} \\
\hline
\vphantom{$1^{{2^3}^4}$} 
$l$ & $a$ & $b$ & $r$\\
\end{tabular}\ .
\label{ternend}
\end{equation}
%
\begin{scheme}[Safeguarded ternary partition]
\label{sts}
\rm
\begin{description}
\itemsep0pt
\item[]
\item[\ref{sts}1.] [Initialize.]
Set $i:=l$, $p:=i+1$, $j:=r$ and $q:=j-1$.
If $v>x_j$, exchange $x_i\leftrightarrow x_j$ and set $p:=i$;
else if $v<x_j$, set $q:=j$.
\item[\ref{sts}2.] [Increase $i$ until $x_i\ge v$.]
Increase $i$ by $1$; then if $x_i<v$, repeat this step.
\item[\ref{sts}3.] [Decrease $j$ until $x_j\le v$.]
Decrease $j$ by $1$; then if $x_j>v$, repeat this step.
\item[\ref{sts}4.] [Exchange.]
(Here $x_j\le v\le x_i$.)
If $i<j$, exchange $x_i\leftrightarrow x_j$; then
if $x_i=v$, exchange $x_i\leftrightarrow x_p$ and increase $p$ by $1$;
if $x_j=v$, exchange $x_j\leftrightarrow x_q$ and decrease $q$ by $1$;
return to \ref{sts}2.
If $i=j$ (so that $x_i=x_j=v$), increase $i$ by $1$ and
decrease $j$ by $1$.
\item[\ref{sts}5.] [Cleanup.]
Set $a:=l+j-p+1$ and $b:=r-q+i-1$.
Exchange $x[l\colon p-1]\leftrightarrow x[p\colon j]$ and
$x[i\colon q]\leftrightarrow x[q+1\colon r]$.
\end{description}
\end{scheme}

Similarly to scheme \ref{sbs}, scheme \ref{sts} makes $n$ or $n+1$
key comparisons, and at most $n/2$ index comparisons at \ref{sts}4.
Let $n_<$, $n_=$, $n_>$ denote the numbers of low, equal and high
keys (here $j-p+1$, $b-a+1$, $q-i+1$).  Step \ref{sts}4 makes
at most $n/2-1$ ``usual'' swaps $x_i\leftrightarrow x_j$, and
$n_=-1$ or $n_=-2$ ``equal'' swaps when $x_i=v$ or $x_j=v$.
Step \ref{sts}5 makes $\min\{p-l,n_<\}+\min\{r-q,n_>\}$ swaps; in
particular, at most $\min\{n_=,n_<+n_>\}$ swaps.
%
\subsection{Single-index controlled ternary partition}
\label{ss:stind1}
Our ternary version of scheme \ref{sbind1} also employs the
arrangements \eqref{ternbeg}--\eqref{ternmid}.
%
\begin{scheme}[Single-index controlled ternary partition]
\label{stind1}
\rm
\begin{description}
\itemsep0pt
\item[]
\item[\ref{stind1}1.] [Initialize.]
Set $i:=l$, $p:=i+1$, $j:=r+1$ and $q:=j-1$.
\item[\ref{stind1}2.] [Increase $i$ until $x_i\ge v$.]
Increase $i$ by $1$; then if $i\le r$ and $x_i<v$, repeat this step.
\item[\ref{stind1}3.] [Decrease $j$ until $x_j\le v$.]
Decrease $j$ by $1$; then if $x_j>v$, repeat this step.
\item[\ref{stind1}4.] [Exchange.]
(Here $x_j\le v\le x_i$.)
If $i<j$, exchange $x_i\leftrightarrow x_j$; then
if $x_i=v$, exchange $x_i\leftrightarrow x_p$ and increase $p$ by $1$;
if $x_j=v$, exchange $x_j\leftrightarrow x_q$ and decrease $q$ by $1$;
return to \ref{stind1}2.
If $i=j$ (so that $x_i=x_j=v$), increase $i$ by $1$ and
decrease $j$ by $1$.
\item[\ref{stind1}5.] [Cleanup.]
Set $a:=l+j-p+1$ and $b:=r-q+i-1$.
Exchange $x[l\colon p-1]\leftrightarrow x[p\colon j]$ and
$x[i\colon q]\leftrightarrow x[q+1\colon r]$.
\end{description}
\end{scheme}

The comparison and swap counts of scheme \ref{stind1} are similar to
those of scheme \ref{sts}; in particular, step \ref{stind1}5 makes
$\min\{p-l,n_<\}+\min\{r-q,n_>\}$ swaps, where $p-l+r-q=n_=$ or
$n_=-1$.  In contrast, a similar scheme of Sedgewick
\cite[Chap.\ 7, quicksort]{sed:ac++3} swaps all the $n_=$ equal keys
in its last step.  More importantly, Sedgewick's scheme needn't produce
{\em true\/} ternary partitions (e.g., for $x=[0,1,0]$ and $v=0$, it
doesn't change the array).
%
\subsection{Double-index controlled ternary partition}
\label{ss:stind2}
We now present our modification of the ternary scheme of
\cite{bemc:esf}, described also in \cite[Prog.\ 1]{bese:fas} and
\cite[Ex.\ 5.2.2--41]{knu:acpIII2}.  It employs the loop invariant
\eqref{ternbeg}, and the cross-over arrangement \eqref{ternmid} with
$j=i-1$ for the swaps leading to the partition \eqref{ternend}.
%
\begin{scheme}[Double-index controlled ternary partition]
\label{stind2}
\rm
\begin{description}
\itemsep0pt
\item[]
\item[\ref{stind2}1.] [Initialize.]
Set $i:=p:=l+1$ and $j:=q:=r$.
\item[\ref{stind2}2.] [Increase $i$ until $x_i>v$.]
If $i\le j$ and $x_i<v$, increase $i$ by $1$ and repeat this step.
If $i\le j$ and $x_i=v$, exchange $x_p\leftrightarrow x_i$, increase
$p$ and $i$ by $1$, and repeat this step.
\item[\ref{stind2}3.] [Decrease $j$ until $x_j<v$.]
If $i<j$ and $x_j>v$, decrease $j$ by $1$ and repeat this step.
If $i<j$ and $x_j=v$, exchange $x_j\leftrightarrow x_q$, decrease
$j$ and $q$ by $1$, and repeat this step.
If $i\ge j$, set $j:=i-1$ and go to \ref{stind2}5.
\item[\ref{stind2}4.] [Exchange.]
Exchange $x_i\leftrightarrow x_j$, increase $i$ by $1$,
decrease $j$ by $1$, and return to \ref{stind2}2.
\item[\ref{stind2}5.] [Cleanup.]
Set $a:=l+i-p$ and $b:=r-q+j$.
Swap $x[l\colon p-1]\leftrightarrow x[p\colon j]$ and
$x[i\colon q]\leftrightarrow x[q+1\colon r]$.
\end{description}
\end{scheme}

Steps \ref{stind2}2 and \ref{stind2}3 make $n_=-1$ swaps, step
\ref{stind2}4 at most $\min\{n_<,n_>\}\le(n-1)/2$ swaps, and step
\ref{stind2}5 takes
$\min\{p-l,n_<\}+\min\{r-q,n_>\}\le\min\{n_=,n_<+n_>\}$ swaps.

Scheme \ref{stind2} makes $n-1$ comparisons, whereas the versions of
\cite[Progs.\ 6 and 7]{bemc:esf}, \cite[\S5]{bese:fas},
\cite[Ex.\ 5.2.2--41]{knu:acpIII2} make one spurious comparison when
$i=j$ at step \ref{stind2}3.  These versions correspond to replacing
step \ref{stind2}3 by
\begin{description}
\itemsep0pt
\item[\ref{stind2}3'.] [Decrease $j$ until $x_j<v$.]
If $i\le j$ and $x_j>v$, decrease $j$ by $1$ and repeat this step.
If $i\le j$ and $x_j=v$, exchange $x_j\leftrightarrow x_q$, decrease
$j$ and $q$ by $1$, and repeat this step.
If $i\ge j$, go to \ref{stind2}5.
\end{description}

Except for making a spurious comparison when $i=j$, step \ref{stind2}3'
acts like \ref{stind2}3: If $i=j$, then, since $x_i>v$ by \ref{stind2}2,
they exit to \ref{stind2}5 with $j=i-1$, whereas if $i>j$, then the
general invariant $i\le j+1$ yields $i=j+1$, and \ref{stind2}3 maintains
this equality.
%
\begin{lemma}
\label{l:BMcmp}
Let\/ $c\in\{0,1\}$ be the number of spurious comparisons made by
scheme\/ {\rm\ref{stind2}} using step\/ {\rm\ref{stind2}3'}.  If
the keys are distinct and in random order, then\/
$\Exp[c|j_v=j]=1-j/(n-1)$ for\/ $0\le j<n$, and\/
$\Exp c=1-\Exp j_v/(n-1)$, where\/ $j_v$ is the number of keys\/
$<v$.  In particular, $\Exp c=1/2$ when the pivot\/ $v$ is the
median-of-$s$ {\rm(}for odd\/ $s\ge1${\rm)} or the ninther
{\rm(}cf.\ \S{\rm\ref{ss:pivsample});} in these cases scheme\/
{\rm\ref{stind2}} with step\/ {\rm\ref{stind2}3'} makes on average\/
$n-1/2$ comparisons.
\end{lemma}
\begin{proof}
For distinct keys, the final $i=a+1$ and $j=a$ at step \ref{stind2}5.
If $c=1$, then $i=j$ and $x_i>v$ at \ref{stind2}3' yield $i=a+1\le n$
and $x_{a+1}>v$.  Conversely, suppose $a<n$ and $x_{a+1}>v$ on input.
If $x_{a+1}$ were compared to $v$ first at \ref{stind2}3' for
$j=a+1>i$, \ref{stind2}3' would set $j=a$ and exit to \ref{stind2}5
(since \ref{stind2}4 would decrease $j$ below $a$) with
$i\le a$, a contradiction; hence $x_{a+1}$ must be compared to $v$
first at \ref{stind2}2 for $i=a+1\le j$, and again at \ref{stind2}3'.
Thus $c=1$ iff $a<n$ and $x_{a+1}>v$ on input.  Consequently, for
$j_v:=a-1=j<n-1$, $\Exp[c|j_v=j]=\Pr[x_{a+1}>v|j_v=j]=(n-1-j)/(n-1)$
since there are $n-1-j$ high keys in random order, and
$\Exp[c|j_v=n-1]=0$; the rest is straighforward.
\qed
\end{proof}
%
\subsection{Lomuto's ternary partition}
\label{ss:stL}
Our ternary extension of scheme \ref{sbL} employs the following
``strict'' version of \eqref{sbLmid}:
\begin{equation}
\begin{tabular}{lrrclr}
\hline
\multicolumn{2}{|c|}{$x=v$} &
\multicolumn{1}{|c|}{$x<v$} &
\multicolumn{1}{|c|}{$x>v$} &
\multicolumn{2}{|c|}{$?$} \\
\hline
\vphantom{$1^{{2^3}^4}$} 
$l$ & $\bar p$ & $p$ & & $i$ & $r$ \\
\end{tabular}
\quad\longrightarrow\quad
\begin{tabular}{lrrr}
\hline
\multicolumn{2}{|c|}{$x=v$} &
\multicolumn{1}{|c|}{$x<v$} &
\multicolumn{1}{|c|}{$x>v$} \\
\hline
\vphantom{$1^{{2^3}^4}$} 
$l$ & $\bar p$ & $p$ & $r$ \\
\end{tabular}\ .
\label{stLmid}
\end{equation}
%
\begin{scheme}[Lomuto's ternary partition]
\label{stL}
\rm
\begin{description}
\itemsep0pt
\item[]
\item[\ref{stL}1.] [Initialize.]
Set $i:=l+1$ and $p:=\bar p:=l$.
\item[\ref{stL}2.] [Check if done.]
If $i>r$, go to \ref{stL}4.
\item[\ref{stL}3.] [Exchange if necessary.]
If $x_i<v$, increase $p$ by $1$ and exchange $x_p\leftrightarrow x_i$.
If $x_i=v$, increase $\bar p$ and $p$ by $1$ and exchange
$x_p\leftrightarrow x_i$ and $x_{\bar p}\leftrightarrow x_p$.
Increase $i$ by $1$ and return to \ref{stL}2.
\item[\ref{stL}4.] [Cleanup.]
Set $a:=l+p-\bar p$ and $b:=p$.  Exchange
$x[l\colon\bar p]\leftrightarrow x[\bar p+1\colon p]$.
\end{description}
\end{scheme}

Scheme \ref{stL} makes $n_<+2(n_=-1)+\min\{n_=,n_<\}$ swaps.
Using the arrangements
\begin{equation}
\begin{tabular}{lrrclr}
\hline
\multicolumn{2}{|c|}{$x<v$} &
\multicolumn{1}{|c|}{$x=v$} &
\multicolumn{1}{|c|}{$x>v$} &
\multicolumn{2}{|c|}{$?$} \\
\hline
\vphantom{$1^{{2^3}^4}$} 
$l$ & $\bar p$ & $p$ & & $i$ & $r$ \\
\end{tabular}
\quad\longrightarrow\quad
\begin{tabular}{lrrr}
\hline
\multicolumn{2}{|c|}{$x<v$} &
\multicolumn{1}{|c|}{$x=v$} &
\multicolumn{1}{|c|}{$x>v$} \\
\hline
\vphantom{$1^{{2^3}^4}$} 
$l$ & $\bar p$ & $p$ & $r$ \\
\end{tabular}\ ,
\label{stLmidalt}
\end{equation}
with obvious modifications, scheme \ref{stL} would make $n_=-1+2n_<$
swaps.
%
\subsection{Comparison of binary and ternary schemes}
\label{ss:compbitern}
When the keys are distinct, the binary schemes \ref{sbs}, \ref{sbind1},
\ref{sbind2}, \ref{sbL} are {\em equivalent\/} to their ternary versions
\ref{sts}, \ref{stind1}, \ref{stind2}, \ref{stL} in the sense that
respective pairs of schemes (e.g., \ref{sbs} and \ref{sts}) produce
identical partitions, making the same sequences of comparisons and swaps.
Hence our results of \S\ref{s:averbi} extend to the ternary schemes by
replacing \ref{sbs}, \ref{sbind1}, \ref{sbind2}, \ref{sbL} with
\ref{sts}, \ref{stind1}, \ref{stind2}, \ref{stL}, respectively.  Since
the overheads of the ternary schemes are relatively small, consisting
mostly of additional tests for equal keys, the ternary schemes should
run almost as fast as their binary counterparts in the case of distinct
keys.

Let us highlight some differences when equal keys occur.  Although
schemes \ref{sbs} and \ref{sts} stop the scanning pointers $i$ and $j$
on the same keys, step \ref{sbs}4 simply swaps each key to the other
side, whereas step \ref{sts}4 additionally swaps equals to the ends.
Schemes \ref{sbind1} and \ref{stind1} behave similarly.  However,
in contrast with scheme \ref{sbind2}, scheme \ref{stind2} never swaps
equals to the other side.  For instance, when all the keys are equal,
scheme \ref{sts} makes $\lfloor n/2-1\rfloor$ usual swaps and
$2\lfloor n/2-1\rfloor$ vacuous swaps, scheme \ref{stind1} makes
$\lfloor(n-1)/2\rfloor$ usual swaps and $2\lfloor(n-1)/2\rfloor$
vacuous swaps, scheme \ref{stind2} makes just $n-1$ vacuous swaps, and
scheme \ref{stL} makes $2(n-1)$ vacuous swaps.
%
\subsection{Preventing vacuous swaps of equal keys}
\label{ss:prevvac}
Steps \ref{stind2}2 and \ref{stind2}3 of scheme \ref{stind2} have two
drawbacks: they make vacuous swaps when $i=p$ and $j=q$, and they
need the tests ``$i\le j$'' and ``$i<j$''.  These drawbacks are
eliminated in the following two-phase scheme, which runs first a
special version of scheme \ref{stind2} that doesn't make vacuous swaps
until it finds two keys $x_i<v<x_j$.  Afterwards no vacuous swaps occur
(because $p<i$, $j<q$) and the pointer tests are unnecessary (since
$x_j>v$ stops the $i$-loop, and $x_{i-1}<v$ stops the $j$-loop).
%
\begin{scheme}[Hybrid ternary partition]
\label{sth}
\rm
\begin{description}
\itemsep0pt
\item[]
\item[\rlap{\ref{sth}1.}\phantom{\ref{sth}11.}] [Initialize.]
Set $i:=l+1$ and $j:=q:=r$.
\item[\rlap{\ref{sth}2.}\phantom{\ref{sth}11.}]
[Increase $i$ until $x_i\ne v$.]
If $i\le j$ and $x_i=v$, increase $i$ by $1$ and repeat this step.
Set $p:=i$.
If $i=j$, set $i:=j+1$ if $x_i<v$, $j:=i-1$ otherwise.
If $i\ge j$, go to \ref{sth}12.
\item[\rlap{\ref{sth}3.}\phantom{\ref{sth}11.}]
[Decrease $j$ until $x_j\ne v$.]
If $i<j$ and $x_j=v$, decrease $j$ by $1$ and repeat this step.
Set $q:=j$.
If $i=j$, set $i:=j+1$ if $x_i<v$, $j:=i-1$ otherwise, and
go to \ref{sth}12.
\item[\rlap{\ref{sth}4.}\phantom{\ref{sth}11.}]
[Decide which steps to skip.]
If $x_i<v$ and $x_j<v$, go to \ref{sth}5.
If $x_i>v$ and $x_j>v$, go to \ref{sth}6.
If $x_i>v$ and $x_j<v$, go to \ref{sth}7.
If $x_i<v$ and $x_j>v$, go to \ref{sth}8.
\item[\rlap{\ref{sth}5.}\phantom{\ref{sth}11.}]
[Increase $i$ until $x_i>v$.]
Increase $i$ by $1$.
If $i<j$ and $x_i<v$, repeat this step.
If $i<j$ and $x_i=v$, exchange $x_p\leftrightarrow x_i$, increase
$p$ by $1$, and repeat this step.
(At this point, $x_j<v$.)
If $i<j$, go to \ref{sth}7.
Set $i:=j+1$ and go to \ref{sth}12.
\item[\rlap{\ref{sth}6.}\phantom{\ref{sth}11.}]
[Decrease $j$ until $x_j<v$.]
Decrease $j$ by $1$.
If $i<j$ and $x_j>v$, repeat this step.
If $i<j$ and $x_j=v$, exchange $x_j\leftrightarrow x_q$, decrease
$q$ by $1$, and repeat this step.
(At this point, $x_i>v$.)
If $i=j$, set $j:=i-1$ and go to \ref{sth}12.
\item[\rlap{\ref{sth}7.}\phantom{\ref{sth}11.}] [Exchange.]
(At this point, $i<j$ and $x_i>v>x_j$.)
Exchange $x_i\leftrightarrow x_j$.
\item[\rlap{\ref{sth}8.}\phantom{\ref{sth}11.}]
[End of first stage.]
(At this point, $x_i<v<x_j$ and $p\le i<j\le q$.)
\item[\rlap{\ref{sth}9.}\phantom{\ref{sth}11.}]
[Increase $i$ until $x_i>v$.]
Increase $i$ by $1$.
If $x_i<v$, repeat this step.
If $x_i=v$, exchange $x_p\leftrightarrow x_i$, increase
$p$ by $1$, and repeat this step.
\item[\rlap{\ref{sth}10.}\phantom{\ref{sth}11.}]
[Decrease $j$ until $x_j<v$.]
Decrease $j$ by $1$.
If $x_j>v$, repeat this step.
If $x_j=v$, exchange $x_j\leftrightarrow x_q$, decrease
$q$ by $1$, and repeat this step.
\item[\rlap{\ref{sth}11.}\phantom{\ref{sth}11.}] [Exchange.]
If $i<j$, exchange $x_i\leftrightarrow x_j$ and
return to \ref{sth}9.
\item[\rlap{\ref{sth}12.}\phantom{\ref{sth}11.}] [Cleanup.]
Set $a:=l+i-p$ and $b:=r-q+j$.
Exchange $x[l\colon p-1]\leftrightarrow x[p\colon j]$ and
$x[i\colon q]\leftrightarrow x[q+1\colon r]$.
\end{description}
\end{scheme}

Scheme \ref{sth} makes $n+1$ comparisons, or just $n-1$ if it finishes
in the first stage before reaching step \ref{sth}9.  The two
extraneous comparisons can be eliminated by keeping the strategy of
scheme \ref{stind2} in the following modification.
%
\begin{scheme}[Extended double-index controlled ternary partition]
\label{stind2h}
\rm
\mbox{}\\
Use scheme \ref{sth} with steps \ref{sth}8 through \ref{sth}11
replaced by the following steps.
\vspace{-4pt}
\begin{description}
\itemsep0pt
\item[\rlap{\ref{sth}8.}\phantom{\ref{sth}11.}] [End of first stage.]
Increase $i$ by $1$ and decrease $j$ by $1$.
\item[\rlap{\ref{sth}9.}\phantom{\ref{sth}11.}]
[Increase $i$ until $x_i>v$.]
If $i\le j$ and $x_i<v$, increase $i$ by $1$ and repeat this step.
If $i\le j$ and $x_i=v$, exchange $x_p\leftrightarrow x_i$, increase
$p$ and $i$ by $1$, and repeat this step.
\item[\ref{sth}10.] [Decrease $j$ until $x_j<v$.]
If $i<j$ and $x_j>v$, decrease $j$ by $1$ and repeat this step.
If $i<j$ and $x_j=v$, exchange $x_j\leftrightarrow x_q$, decrease
$j$ and $q$ by $1$, and repeat this step.
If $i\ge j$, set $j:=i-1$ and go to \ref{sth}12.
\item[\ref{sth}11.] [Exchange.]
Exchange $x_i\leftrightarrow x_j$, increase $i$ by $1$,
decrease $j$ by $1$, and return to \ref{sth}9.
\end{description}
\end{scheme}

Schemes \ref{sth} and \ref{stind2h} are {\em equivalent\/} in the
sense of producing identical partitions via the same sequences of
swaps.  Further, barring vacuous swaps, scheme \ref{stind2} is
equivalent to schemes \ref{sth} and \ref{stind2h} in the following
cases: (a) all keys are equal; (b) $x_r\ne v$ (e.g., the keys are
distinct); (c) there is at least one high key $>v$.  In the remaining
{\em degenerate\/} case where the keys aren't equal, $x_r=v$ and there
are no high keys, scheme \ref{stind2} produces $i=r+1$ and $j=r$ on
the first pass, whereas step \ref{sth}3 finds $j<r$, and either
\ref{sth}3 or \ref{sth}5 produce $i=j+1$ (i.e., scheme \ref{stind2}
swaps $r-j$ more equal keys to the left end).

If the first stage of schemes \ref{sth} and \ref{stind2h} is implemented
by a more straightforward adaptation of scheme \ref{stind2}, we obtain
the following variants.
%
\begin{scheme}[Alternative hybrid ternary partition]
\label{sthalt}
\rm
\mbox{}\\
Use scheme \ref{sth} with steps \ref{sth}2 through \ref{sth}6
replaced by the following steps.
\vspace{-4pt}
\begin{description}
\itemsep0pt
\item[\rlap{\ref{sth}2.}\phantom{\ref{sth}11.}]
[Increase $i$ until $x_i\ne v$.]
If $i\le j$ and $x_i=v$, increase $i$ by $1$ and repeat this step.
Set $p:=i$.  If $i\le j$ and $x_i<v$, increase $i$ by $1$ and go to
\ref{sth}3; otherwise go to \ref{sth}4.
\item[\rlap{\ref{sth}3.}\phantom{\ref{sth}11.}]
[Increase $i$ until $x_i>v$.]
If $i\le j$ and $x_i<v$, increase $i$ by $1$ and repeat this step.
If $i\le j$ and $x_i=v$, exchange $x_p\leftrightarrow x_i$, increase
$p$ and $i$ by $1$, and repeat this step.
\item[\rlap{\ref{sth}4.}\phantom{\ref{sth}11.}]
[Decrease $j$ until $x_j\ne v$.]
If $i<j$ and $x_j=v$, decrease $j$ by $1$ and repeat this step.
Set $q:=j$.
If $i<j$ and $x_j>v$, decrease $j$ by $1$ and go to \ref{sth}5.
If $i<j$ and $x_j<v$, go to \ref{sth}7.
Set $j:=i-1$ and go to \ref{sth}12.
\item[\rlap{\ref{sth}5.}\phantom{\ref{sth}11.}]
[Decrease $j$ until $x_j<v$.]
If $i<j$ and $x_j>v$, decrease $j$ by $1$ and repeat this step.
If $i<j$ and $x_j=v$, exchange $x_j\leftrightarrow x_q$, decrease
$j$ and $q$ by $1$, and repeat this step.
If $i\ge j$, set $j:=i-1$ and go to \ref{sth}12.
\end{description}
\end{scheme}
%
\begin{scheme}[Two-stage double-index controlled ternary partition]
\label{stind2halt}
\rm
\mbox{}\\
Use scheme \ref{sth} with steps \ref{sth}2 through \ref{sth}6
replaced by steps \ref{sth}2 through \ref{sth}5 of scheme \ref{sthalt},
and steps \ref{sth}8 through \ref{sth}11 replaced by steps \ref{sth}8
through \ref{sth}11 of scheme \ref{stind2h}.
\end{scheme}

In other words, scheme \ref{stind2halt} is obtained from scheme
\ref{stind2} by using special versions of steps \ref{stind2}2 and
\ref{stind2}3 on the first pass, with each step split into two
substeps to avoid vacuous swaps.

Except for avoiding vacuous swaps, schemes \ref{sthalt} and
\ref{stind2halt} are equivalent to scheme \ref{stind2}.  Hence schemes
\ref{stind2}, \ref{sth}, \ref{stind2h}, \ref{sthalt} and
\ref{stind2halt} are equivalent except for the degenerate case
discussed after scheme \ref{stind2h}; in this case, schemes \ref{sth}
and \ref{stind2h} swap fewer equal keys than schemes \ref{sthalt} and
\ref{stind2halt}.  Another significant difference between schemes
\ref{sth} and \ref{sthalt} is that scheme \ref{sth}
may be quicker in reaching the second stage where the tests
``$i\le j$'' and ``$i<j$'' aren't needed.  (In fact scheme \ref{sth}
reaches step \ref{sth}8 faster than scheme \ref{sthalt} iff
$x_i<v<x_j$ occurs at step \ref{sth}4 of scheme \ref{sth}; in the
remaining three cases of \ref{sth}4 both schemes act equivalently.)
%
\subsection{Using sample elements in tripartitioning}
\label{s:sampsenttri}
In parallel with \S\ref{s:sampsent}, we now show how to tune the
ternary schemes when the pivot $v$ is selected as the $(\hat p+1)$th
element in a sample of size $s$, assuming $0\le\hat p<s\le n$ and
$\hat q:=s-1-\hat p>0$.

First, suppose that after pivot selection, we have the following
arrangement:
\begin{equation}
\begin{tabular}{lllrrr}
\hline
\multicolumn{1}{|c|}{$x<v$} &
\multicolumn{1}{|c|}{$x=v$} &
\multicolumn{2}{|c|}{?} &
\multicolumn{1}{|c|}{$x=v$} &
\multicolumn{1}{|c|}{$x>v$} \\
\hline
\vphantom{$1^{{2^3}^4}$} 
$l$ & $\bar l$ & $p$ & $q$ & $\bar r$ & $r$\\
\end{tabular}\ ,
\label{ternbegsample}
\end{equation}
with $p:=l+\hat p+1$, $q:=r-\hat q$; then we only need to partition the
array $x[p-1\colon q]$ of size $\bar n:=n-s+1$.  The ternary schemes
are modified as follows.

In step \ref{sts}1 of scheme \ref{sts}, set $i:=p-1$ and $j:=q+1$; in
step \ref{sts}5 replace $l$, $r$ by $\bar l$, $\bar r$.  The same
scheme results from scheme \ref{stind1} after analogous changes and
omitting the test ``$i\le r$'' in \ref{stind1}2.  Similarly, in step
\ref{stind2}1 of scheme \ref{stind2}, set $i:=p$ and $j:=q$; in step
\ref{stind2}5 replace $l$, $r$ by $\bar l$, $\bar r$.  Steps \ref{sth}1
and \ref{sth}11 of schemes \ref{sth} through \ref{stind2halt} are
modified in the same way.  Finally, in step \ref{stL}1 of scheme
\ref{stL} set $i:=p$ and $p:=\bar p:=i-1$; in step \ref{stL}2 replace
$r$ by $q$; in step \ref{stL}4 set $a:=\bar l+p-\bar p$,
$b:=p-q+\bar r$ and exchange
$x[\bar l\colon\bar p]\leftrightarrow x[\bar p+1\colon p]$ and
$x[p+1\colon q]\leftrightarrow x[q+1\colon\bar r]$.

When the keys are distinct, we have $\bar l=l+\hat p$, $p=\bar l+1$
and $q=\bar r=r-\hat q$ in \eqref{ternbegsample}, so that schemes
\ref{sts}, \ref{stind1}, \ref{stind2}, \ref{stL} are equivalent to
schemes \ref{sbs}, \ref{sbind1}, \ref{sbind2}, \ref{sbL} as modified
in \S\ref{s:sampsent} (where $p$, $q$ correspond to the current
$\hat p$, $\hat q$).

For the median-of-3 selection ($\hat p=\hat q=1$, $p=l+2$,
$q=r-1$), we may rearrange the sample keys $x_l$, $x_{l+1}$, $x_r$
and find $\bar l$, $\bar r$ according to Figure \ref{fig:med3}.
%
\begin{figure}
\footnotesize
\begin{center}
\setlength{\unitlength}{1.0mm}%
\begin{picture}(161,37)(0,17.5)
\put(0,40){\begin{picture}(161,20)
\put(80,11){\circle{8}}
\put(80,11){\makebox(0,0){$a:c$}}
\put(34,2){\line(6,1){42}}
\put(80,4){\line(0,1){3}}
\put(126,2){\line(-6,1){42}}
\put(55,8){\makebox(0,0){$<$}}
\put(82,5){\makebox(0,0)[l]{$=$}}
\put(105,8){\makebox(0,0){$>$}}
\put(30,0){\circle{8}}
\put(30,0){\makebox(0,0){$b:a$}}
\put(80,0){\circle{8}}
\put(80,0){\makebox(0,0){$a:b$}}
\put(130,0){\circle{8}}
\put(130,0){\makebox(0,0){$b:a$}}
\end{picture}}
\put(12,32.5){\line(3,1){14.5}}
\put(30,32.5){\line(0,1){3.5}}
\put(42.5,32.5){\line(-2,1){9.5}}
\put(61,32.5){\line(3,1){15.5}}
\put(80,32.5){\line(0,1){3.5}}
\put(99,32.5){\line(-3,1){15.5}}
\put(117.5,32.5){\line(2,1){9.5}}
\put(130,32.5){\line(0,1){3.5}}
\put(148,32.5){\line(-3,1){14.5}}
\put(4,27.5){\begin{picture}(152,5)
\put(0,0){\framebox(16,5){$b<a<c$}}
\put(19,0){\framebox(16,5){$a=b<c$}}
\put(42,2.5){\circle{8}}
\put(42,2.5){\makebox(0,0){$b:c$}}
\put(49,0){\framebox(16,5){$a=c<b$}}
\put(68,0){\framebox(16,5){$a=b=c$}}
\put(87,0){\framebox(16,5){$b<a=c$}}
\put(110,2.5){\circle{8}}
\put(110,2.5){\makebox(0,0){$b:c$}}
\put(117,0){\framebox(16,5){$c<b=a$}}
\put(136,0){\framebox(16,5){$c<a<b$}}
\end{picture}}
\put(0,17.5){\begin{picture}(161,5)
\put(19,0){\framebox(16,5){$a<b<c$}}
\put(38,0){\framebox(16,5){$a<b=c$}}
\put(57,0){\framebox(16,5){$a<c<b$}}
\put(87,0){\framebox(16,5){$b<c<a$}}
\put(106,0){\framebox(16,5){$c=b<a$}}
\put(125,0){\framebox(16,5){$c<b<a$}}
\end{picture}}
\put(27,22.5){\line(3,1){15.5}}
\put(46,22.5){\line(0,1){3.5}}
\put(65,22.5){\line(-3,1){15.5}}
\put(95,22.5){\line(3,1){15.5}}
\put(114,22.5){\line(0,1){3.5}}
\put(133,22.5){\line(-3,1){15.5}}
\end{picture}
\end{center}
\caption{Decision tree for median of three}
\label{fig:med3}
\end{figure}
(For simplicity, as with Fig.\ \ref{fig:med3pm}, the left subtree
may be used after exchanging $a\leftrightarrow c$ when $a>c$.)

As in \S\ref{s:sampsent}, even if pivot selection doesn't rearrange the
array except for placing the pivot in $x_l$, scheme \ref{sts} may be
simplified by replacing step \ref{sts}1 with step \ref{stind1}1; the
same scheme is obtained from scheme \ref{stind1} by omitting the test
``$i\le r$'' in \ref{stind1}2.
%
\section{Experimental results}
\label{s:exp}
%
\subsection{Implemented algorithms}
\label{ss:impl}
We now sketch the algorithms used in our experiments, starting with
a nonrecursive version of quickselect that employs a random pivot and
one of the ternary schemes of \S\ref{s:terpart}.
%
\begin{algorithm}[{{\rm{\sc Quickselect}$(x,n,k)$ for selecting the
$k$th smallest of $x[1\colon n{]}$}}]
\label{alg:qsel}
\rm
\mbox{}
\medbreak\noindent{\bf Step 1} ({\em Initialize\/}).
Set $l:=1$ and $r:=n$.
\medbreak\noindent{\bf Step 2} ({\em Handle small file\/}).
If $l<r$, go to Step 3.  If $l>r$, set $k_-:=r+1$ and $k_+:=l-1$.
If $l=r$, set $k_-:=k_+:=k$.  Return.
\medbreak\noindent{\bf Step 3} ({\em Select pivot\/}).
Pick a random integer $i\in[l,r]$, swap $x_l\leftrightarrow x_i$
and set $v:=x_l$.
\medbreak\noindent{\bf Step 4} ({\em Partition\/}).
Partition the array $x[l\colon r]$ to produce the arrangement
\eqref{ternend}.
\medbreak\noindent{\bf Step 5} ({\em Update bounds\/}).
If $a\le k$, set $l:=b+1$.  If $k\le b$, set $r:=a-1$.  Go to
Step 2.
\end{algorithm}

Steps 2 and 5 ensure that on exit
$x[1\colon k_--1]< x[k_-\colon k_+]<x[k_++1\colon n]$,
$k_-\le k\le k_+$.

The median-of-3 version works as follows.  If $l=r-1$ at Step 2, we
swap $x_l\leftrightarrow x_r$ if $x_l>x_r$, set $k_-:=l$ and $k_+:=r$
if $x_l=x_r$, $k_-:=k_+:=k$ otherwise, and return.  At Step 3, we
swap $x_{l+1}$, $x_r$ with random keys in $x[l+1\colon r]$ and
$x[l+2\colon r]$, respectively.  After sorting the sample keys $x_l$,
$x_{l+1}$, $x_r$ and finding $\bar l$, $\bar r$ for
\eqref{ternbegsample} according to Fig.\ \ref{fig:med3}, we set
$v:=x_{l+1}$.  Then Step 4 uses one of the modified ternary schemes of
\S\ref{s:sampsenttri}.

When a binary scheme is employed, we omit $k_-$ and $k_+$, use
Fig.\ \ref{fig:med3pm} instead of Fig.\ \ref{fig:med3}, and the
modified schemes of \S\ref{s:sampsent} with $\bar l:=l+1$,
$\bar r:=r-1$ for the median-of-3.

Our implementations of {\sc Quickselect} were programmed in Fortran 77
and run on a notebook PC (Pentium 4M 2 GHz, 768 MB RAM) under MS
Windows XP.  We used a double precision input array $x[1\colon n]$,
in-line comparisons and swaps; future work should test tuned comparison
and swap functions for other data types (cf.\ \cite{bemc:esf}).

%
\subsection{Testing examples}
\label{ss:examp}
We used minor modifications of the input sequences of \cite{val:iss},
defined as follows:
\begin{description}
\itemsep0pt
\item[random]
A random permutation of the integers $1$ through $n$.
\item[mod-$m$]
A random permutation of the sequence $i\bmod m$, $i=1\colon n$,
called {\em binary\/} ({\em ternary\/}, {\em quadrary\/},
{\em quintary\/}) when $m=2$ ($3$, $4$, $5$, respectively).
\item[sorted]
The integers $1$ through $n$ in increasing order.
\item[rotated]
A sorted sequence rotated left once; i.e., $(2,3,\ldots,n,1)$.
\item[organpipe]
The integers $(1,2,\ldots,n/2,n/2,\ldots,2,1)$.
\item[m3killer]
Musser's ``median-of-3 killer'' sequence with $n=4j$ and $k=n/2$:
$$
\left(\begin{array}{ccccccccccccc}
1&  2 & 3&  4 & \ldots&  k-2& k-1& k& k+1& \ldots& 2k-2& 2k-1& 2k\\
1& k+1& 3& k+3& \ldots& 2k-3& k-1& 2&  4 & \ldots& 2k-2& 2k-1& 2k
\end{array}\right).
$$
\item[twofaced]
Obtained by randomly permuting the
elements of an m3killer sequence in positions $4\lfloor\log_2n\rfloor$
through $n/2-1$ and $n/2+4\lfloor\log_2n\rfloor-1$ through $n-2$.
\end{description}
For each input sequence, its (lower) median element was selected
for $k:=\lceil n/2\rceil$.

These input sequences were designed to test the performance of
selection algorithms under a range of conditions.
In particular, the binary sequences represent inputs containing
many duplicates \cite{sed:qek}.  The rotated and organpipe sequences
are difficult for many implementations of quickselect.  The m3killer
and twofaced sequences are hard for implementations with median-of-3
pivots (their original versions \cite{mus:iss} were modified to become
difficult when the middle element comes from position $k$ instead of
$k+1$).
%
\subsection{Computational results}
\label{ss:result}
We varied the input size $n$ from $50{,}000$ to $16{,}000{,}000$.  For
the random, mod-$m$ and twofaced sequences, for each input size,
20 instances were randomly generated; for the deterministic
sequences, 20 runs were made to measure the solution time.

Table \ref{tab:dsel02} summarizes the performance of four schemes used
in {\sc Quickselect} with median-of-3.
%
\begin{table}
\caption{Performance of schemes \ref{sbs}, \ref{sts}, \ref{stind2},
\ref{sth} with median-of-3.}
\label{tab:dsel02}
\footnotesize
\begin{center}
\renewcommand{\arraystretch}{0.98745}
\begin{tabular}{clrrrrrrrrrrr}
\hline
\multicolumn{1}{c}{Scheme}&
Sequence &\multicolumn{1}{c}{Size}
&\multicolumn{3}{c}{Time $[{\rm msec}]$%
\vphantom{$1^{2^3}$}} 
&\multicolumn{3}{c}{Comparisons $[n]$}
&\multicolumn{1}{c}{$P_{\rm avg}$}
&\multicolumn{1}{c}{$S_{\rm avg}$}
&\multicolumn{1}{c}{$S_{\rm avg}^0$}
&\multicolumn{1}{c}{$S_{\rm avg}$}\\
& &\multicolumn{1}{c}{$n$}
&\multicolumn{1}{c}{avg}&\multicolumn{1}{c}{max}&\multicolumn{1}{c}{min}
&\multicolumn{1}{c}{avg}&\multicolumn{1}{c}{max}&\multicolumn{1}{c}{min}
&\multicolumn{1}{c}{$[\ln n]$}
&\multicolumn{1}{c}{$[n]$}
&\multicolumn{1}{c}{$[n]$}
&\multicolumn{1}{c}{$[C_{\rm avg}]$}\\
\hline
\ref{sbs}  &
random     &   8M
&  252&  360&  170& 2.59& 3.96& 1.78& 1.64& 0.55& 0.00& 0.21\\
&
           &  16M
&  494&  641&  371& 2.57& 3.46& 1.93& 1.57& 0.53& 0.00& 0.21\\
&
organpipe  &   8M
&  173&  250&  111& 2.64& 4.10& 1.77& 1.53& 0.57& 0.00& 0.22\\
&
           &  16M
&  355&  460&  270& 2.61& 3.49& 1.94& 1.62& 0.60& 0.00& 0.23\\
&
binary     &   8M
&  254&  271&  250& 2.73& 2.92& 2.68& 1.86& 1.00& 0.00& 0.37\\
&
           &  16M
&  506&  521&  500& 2.70& 2.79& 2.68& 1.87& 1.00& 0.00& 0.37\\
&
ternary    &   8M
&  246&  321&  171& 2.44& 3.27& 1.75& 1.33& 0.82& 0.00& 0.34\\
&
           &  16M
&  452&  620&  360& 2.22& 3.11& 1.75& 1.29& 0.76& 0.00& 0.34\\
&
quadrary   &   8M
&  277&  340&  230& 2.78& 3.44& 2.26& 1.83& 0.86& 0.00& 0.31\\
&
           &  16M
&  537&  671&  460& 2.65& 3.37& 2.26& 1.85& 0.84& 0.00& 0.32\\
&
quintary   &   8M
&  231&  350&  180& 2.31& 3.56& 1.85& 1.34& 0.69& 0.00& 0.30\\
&
           &  16M
&  486&  671&  330& 2.44& 3.49& 1.67& 1.36& 0.71& 0.00& 0.29\\
\ref{sts} &
random     &   8M
&  284&  391&  201& 2.59& 3.96& 1.78& 1.64& 0.55& 0.00& 0.21\\
&
           &  16M
&  550&  711&  411& 2.57& 3.46& 1.93& 1.57& 0.53& 0.00& 0.21\\
&
organpipe  &   8M
&  232&  321&  120& 2.73& 5.27& 1.84& 1.54& 0.57& 0.00& 0.21\\
&
           &  16M
&  421&  571&  320& 2.92& 4.62& 1.90& 1.58& 0.59& 0.00& 0.20\\
&
binary     &   8M
&  205&  231&  170& 1.28& 1.50& 1.00& 0.10& 1.41& 0.61& 1.11\\
&
           &  16M
&  381&  471&  350& 1.13& 1.50& 1.00& 0.08& 1.19& 0.46& 1.06\\
&
ternary    &   8M
&  259&  281&  240& 1.47& 2.00& 1.00& 0.12& 1.37& 0.37& 0.93\\
&
           &  16M
&  505&  590&  480& 1.37& 2.00& 1.00& 0.10& 1.25& 0.28& 0.91\\
&
quadrary   &   8M
&  262&  331&  210& 1.60& 2.50& 1.00& 0.12& 1.33& 0.28& 0.83\\
&
           &  16M
&  559&  661&  410& 1.66& 2.25& 1.00& 0.13& 1.35& 0.31& 0.81\\
&
quintary   &   8M
&  283&  370&  210& 1.52& 2.40& 1.00& 0.13& 1.14& 0.14& 0.75\\
&
           &  16M
&  582&  731&  420& 1.55& 2.40& 1.00& 0.14& 1.13& 0.14& 0.73\\
\ref{stind2} &
random     &   8M
&  301&  411&  210& 2.59& 3.96& 1.78& 1.64& 0.55& 0.00& 0.21\\
&
           &  16M
&  587&  761&  430& 2.57& 3.46& 1.93& 1.57& 0.53& 0.00& 0.21\\
&
organpipe  &   8M
&  186&  250&  110& 2.88& 4.20& 1.91& 1.55& 0.61& 0.00& 0.21\\
&
           &  16M
&  378&  511&  270& 2.77& 3.93& 1.97& 1.59& 0.59& 0.00& 0.21\\
&
binary     &   8M
&  293&  331&  250& 1.27& 1.50& 1.00& 0.10& 1.27& 0.27& 1.00\\
&
           &  16M
&  549&  671&  500& 1.12& 1.50& 1.00& 0.08& 1.12& 0.13& 1.00\\
&
ternary    &   8M
&  340&  420&  250& 1.47& 2.00& 1.00& 0.12& 1.21& 0.10& 0.82\\
&
           &  16M
&  646&  811&  501& 1.53& 2.00& 1.00& 0.11& 1.26& 0.10& 0.82\\
&
quadrary   &   8M
&  311&  450&  220& 1.42& 2.25& 1.00& 0.12& 1.02& 0.07& 0.72\\
&
           &  16M
&  665&  972&  440& 1.55& 2.50& 1.00& 0.13& 1.13& 0.09& 0.73\\
&
quintary   &   8M
&  319&  451&  220& 1.47& 2.00& 1.00& 0.13& 0.96& 0.07& 0.65\\
&
           &  16M
&  644& 1021&  440& 1.61& 2.80& 1.00& 0.13& 0.97& 0.04& 0.60\\
\ref{sth} &
random     &   8M
&  275&  381&  190& 2.59& 3.96& 1.78& 1.64& 0.55& 0.00& 0.21\\
&
           &  16M
&  536&  681&  391& 2.57& 3.46& 1.93& 1.57& 0.53& 0.00& 0.21\\
&
organpipe  &   8M
&  183&  240&  110& 2.88& 4.20& 1.91& 1.55& 0.61& 0.00& 0.21\\
&
           &  16M
&  357&  461&  260& 2.77& 3.93& 1.97& 1.59& 0.59& 0.00& 0.21\\
&
binary     &   8M
&  245&  261&  230& 1.27& 1.50& 1.00& 0.10& 1.00& 0.00& 0.78\\
&
           &  16M
&  500&  530&  480& 1.12& 1.50& 1.00& 0.08& 1.00& 0.00& 0.89\\
&
ternary    &   8M
&  323&  391&  230& 1.47& 2.00& 1.00& 0.12& 1.11& 0.00& 0.76\\
&
           &  16M
&  620&  761&  470& 1.53& 2.00& 1.00& 0.11& 1.16& 0.00& 0.76\\
&
quadrary   &   8M
&  292&  440&  200& 1.43& 2.25& 1.00& 0.12& 0.95& 0.00& 0.66\\
&
           &  16M
&  630&  922&  420& 1.55& 2.50& 1.00& 0.13& 1.04& 0.00& 0.67\\
&
quintary   &   8M
&  297&  431&  200& 1.47& 2.00& 1.00& 0.13& 0.89& 0.00& 0.60\\
&
           &  16M
&  614& 1042&  411& 1.61& 2.80& 1.00& 0.13& 0.93& 0.00& 0.58\\
\hline
\end{tabular}
\end{center}
\end{table}
The average, maximum and minimum solution times are in milliseconds
(in general, they grow linearly with $n$, and can't be measured
accurately for small inputs; hence only large inputs are included,
with $1{\rm M}:=10^6$).  The comparison counts are in multiples of $n$;
e.g., column seven gives $C_{\rm avg}/n$, where $C_{\rm avg}$ is the
average number of comparisons made over all instances.  Further,
$P_{\rm avg}$ is the average number of partitions in units of $\ln n$,
$S_{\rm avg}$ and $S_{\rm avg}^0$ are the average numbers of all swaps
and of vacuous swaps in units of $n$, and the final column gives the
average number of swaps per comparison.
Note that for random inputs with distinct keys, quickselect with
median-of-3 takes on average $2.75n+o(n)$ comparisons and
$\frac{12}{7}\ln n+o(n)$ partitions \cite{gru:mvh,kimapr:ahf}, and
thus about $0.55n$ swaps when there are $1/5$ swaps per comparison;
e.g., for schemes \ref{sbs}, \ref{sts} and \ref{stind2}.

For each scheme (and others not included in Tab.\ \ref{tab:dsel02}),
the results for the twofaced and m3killer inputs were similar to those
for the random and organpipe inputs, respectively.  The sorted and
rotated inputs were solved about twice faster than the random inputs.

Recall that in tuned versions, scheme \ref{sbind1} coincides with
\ref{sbs} and scheme \ref{stind1} with \ref{sts}.

The run times of schemes \ref{sbind2} and \ref{stind2h} were similar
to those of schemes \ref{sbs} and \ref{sth}, respectively; in other
words, the inclusion of pointer tests in the key comparison loops
didn't result in significant slowdowns.  Also their comparison and
swap counts were similar.

Due to additional tests for equal keys, the ternary schemes were
slower than their binary counterparts on the inputs with distinct keys.
Yet the slowdowns were quite mild (e.g., about ten percent for scheme
\ref{sts} vs.\ \ref{sbs}) and could be considered a fair price for
being able to identify all keys equal to the selected one.  On the
inputs with multiple equal keys, the numbers of comparisons made by
the binary schemes \ref{sbs} and \ref{sbind2} were similar to those
made on the random inputs, but the numbers of swaps increased up
to $n$.  In contrast, the ternary schemes \ref{sts} and \ref{stind2}
took significantly fewer comparisons and more swaps.  Scheme \ref{sts}
produced the largest numbers of swaps, but was still faster than
schemes \ref{stind2} and \ref{stind2h}, whereas scheme \ref{stind2h}
was noticeably faster than scheme \ref{stind2} due to the elimination
of vacuous swaps.

On the inputs with distinct keys, Lomuto's scheme \ref{sbL} was about
sixty percent slower than scheme \ref{sbs}, making about half as many
swaps as comparisons (cf.\ \S\S\ref{ss:pivsamplefixed} and
\ref{s:sampsent}).  On the inputs with multiple equal keys, scheme
\ref{sbL} was really bad: once the current array $x[l\colon r]$
contains only keys equal to the $k$th smallest, each partition removes
two keys, so the running time may be quadratic in the number of equal
keys.  For instance, on a binary input with $k=n/2$, at least
$n(n+20)/16-2$ comparisons are used (if the first $v=1$, we get
$l=1$, $r=k$, and then $l$ increases by $2$ while $r=k$; otherwise the
cost is greater).

Our results were similar while using the classic random pivot instead
of the median-of-3.  Then, for random inputs with distinct keys,
quickselect takes on average $2(1+\ln 2)n+o(n)$ comparisons
\cite[Ex.\ 5.2.2--32]{knu:acpIII2}, and thus about $0.564n$ swaps when
there are $1/6$ swaps per comparison.  Hence, not suprisingly, the
running times and comparison counts on the inputs with distinct keys
increased by between $14$ and $20$ percent, but all the schemes had
essentially the same relative merits and drawbacks as in the
median-of-3 case above.


%
\footnotesize
\newcommand{\noopsort}[1]{} \newcommand{\printfirst}[2]{#1}
  \newcommand{\singleletter}[1]{#1} \newcommand{\switchargs}[2]{#2#1}
\ifx\undefined\bysame
\newcommand{\bysame}{\leavevmode\hbox to3em{\hrulefill}\,}
\fi

\normalsize
%
\end{document}